%% file: main.tex
\documentclass[10pt,twocolumn,letterpaper]{article}

\usepackage{iccv}              

\usepackage[T1]{fontenc}
\usepackage{caption}
\usepackage[labelformat=empty]{subcaption}
\usepackage{bm}
\usepackage{xcolor}
\usepackage{upgreek}
\usepackage{multirow}
\usepackage{colortbl}
\usepackage{pifont}
\usepackage{placeins}
\usepackage{nicefrac}
\usepackage{microtype}
\usepackage{tikz}
\usepackage[symbol]{footmisc}

\usepackage[accsupp]{axessibility}

\newcommand{\xmark}{\ding{55}}


\definecolor{iccvblue}{rgb}{0.21,0.49,0.74}
\usepackage[pagebackref,breaklinks,colorlinks,allcolors=iccvblue]{hyperref}

\def\paperID{13643}
\def\confName{ICCV}
\def\confYear{2025}

\title{AAA-Gaussians: Anti-Aliased and Artifact-Free 3D Gaussian Rendering}

\author{Michael Steiner\footnotemark[1]\hspace{0.4em}\textsuperscript{1} \qquad Thomas Köhler\footnotemark[1]\hspace{0.4em}\textsuperscript{1} \qquad Lukas Radl\textsuperscript{1}\\ Felix Windisch\textsuperscript{1} \qquad Dieter Schmalstieg\textsuperscript{1,2} \qquad Markus Steinberger\textsuperscript{1,3}\\
\textsuperscript{1}Graz University of Technology \qquad \textsuperscript{2}University of Stuttgart
\qquad \textsuperscript{3}Huawei Technologies
}

\begin{document}
\include{commands}
\twocolumn[{
\maketitle
\begin{center}
    \centering
    \captionsetup{type=figure}
    \includegraphics[width=.93\textwidth]{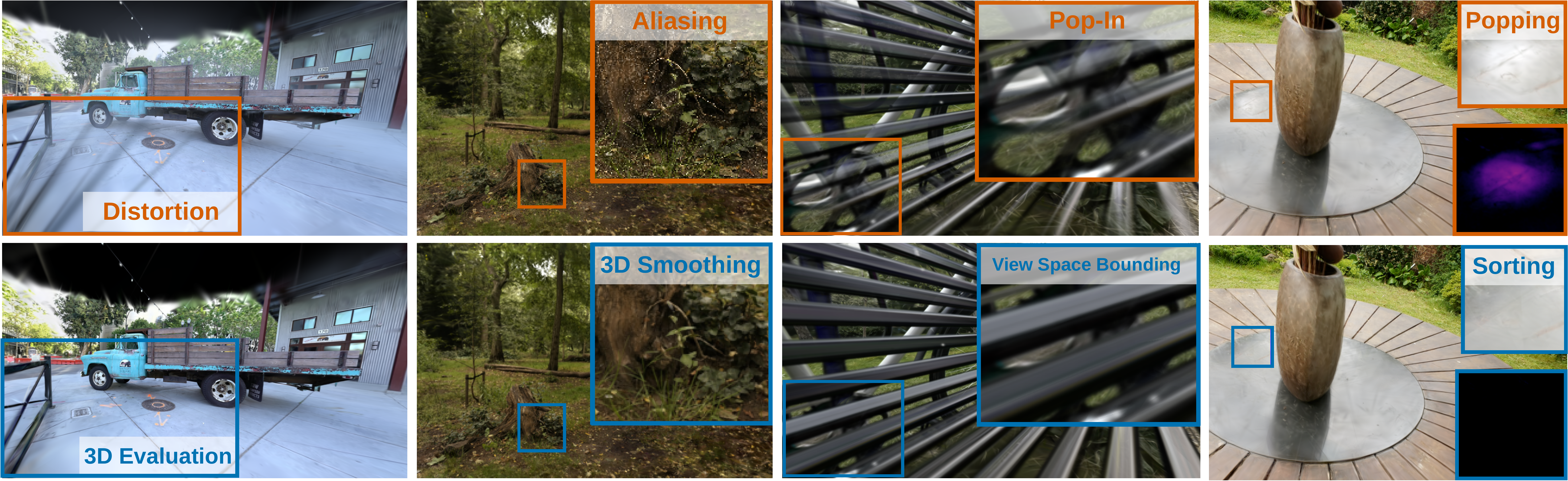}
    \captionof{figure}{
        (Top row) 3DGS rasterization approaches encounter artifacts in out-of-distribution camera settings: (1) Distortions from 2D splat approximations in large field-of-view renderings. (2) 3D evaluation specific aliasing artifacts when zooming out. (3) Incorrect culling results in screen space when the camera is close to objects. (4) Popping due to depth simplifications and global sorting. (Bottom row) Our method addresses these issues with: (1) 3D Gaussian evaluation, (2) a correct aliasing filter, adapted specifically to Gaussian evaluation in 3D, (3) accurate and robust bounding, and (4) efficient 3D culling integrated into hierarchical sorting.
    }
    \label{fig:teaser}
\end{center}%
}]
\footnotetext[1]{Both authors contributed equally to this work}
\begin{abstract}

Although 3D Gaussian Splatting (3DGS) has revolutionized 3D reconstruction, it still faces challenges such as aliasing, projection artifacts, and view inconsistencies, primarily due to the simplification of treating splats as 2D entities. We argue that incorporating full 3D evaluation of Gaussians throughout the 3DGS 
pipeline can effectively address these issues while preserving rasterization efficiency. Specifically, we introduce an adaptive 3D smoothing filter to mitigate aliasing and present a stable view-space bounding method that eliminates popping artifacts when Gaussians extend beyond the view frustum. Furthermore, we promote tile-based culling to 3D with screen-space planes, accelerating rendering and reducing sorting costs for hierarchical rasterization. Our method achieves state-of-the-art quality on in-distribution evaluation sets and significantly outperforms other approaches for out-of-distribution views. Our qualitative evaluations further demonstrate the effective removal of aliasing, distortions, and popping artifacts, ensuring real-time, artifact-free rendering.

\end{abstract}

\input{1_intro}
\input{2_related_work}

\input{3_method}
\input{4_evaluation}
\input{5_conclusion}

\FloatBarrier
{
    \small
    \bibliographystyle{ieeenat_fullname}
    \bibliography{main}
}

\clearpage
\appendix
\input{9_appendix}

\end{document}

%% file: commands.tex
\newcommand{\ifcommentsenabled}[1]{#1}

\definecolor{tab_color}{HTML}{D55E00}

\definecolor{prim_color}{HTML}{0072B2}
\definecolor{second_color}{HTML}{D55E00}
\definecolor{tert_color}{HTML}{009E73}

\definecolor{edited_color}{rgb}{.7,.1,.1}
\newcommand{\new}[1]{#1} 

\definecolor{revised_color}{rgb}{.1,.1,.7}
\newcommand{\revised}[2]{#1} 

%% file: 1_intro.tex
\section{Introduction}
\label{sec:intro}

3D Gaussian Splatting (3DGS)~\cite{kerbl3Dgaussians} has recently revolutionized inverse rendering by enabling fast, differentiable rasterization of 3D Gaussian point clouds.
While 2D splat evaluation is highly efficient, the projection from 3D to 2D Gaussians remains an approximation, introducing artifacts that become particularly noticeable under non-standard camera settings, such as a wide field-of-view in virtual reality rendering.
Additionally, 3DGS further approximates the rendering of 2D splats by assuming them to be parallel to the current view-plane, leading to blend order inconsistencies and popping artifacts under simple camera rotations~\cite{radl2024stopthepop}.

A natural solution to overcome these limitations is to render 3D Gaussians via ray tracing. However, this approach introduces significant computational overhead, requires additional acceleration structures, and is generally impractical for training due to the cost of frequent data structure updates~\cite{loccoz20243dgrt}.

Several works attempt to bridge the gap between 2D splatting and 3D ray tracing by first bounding 3D Gaussians in screen space and then computing their contributions per ray in 3D~\cite{radl2024stopthepop,yu2024gof,hahlbohm2024htgs,wu2024_3dgut,talegaonkar2025volumetric}.
While these hybrid 2D/3D approaches address certain limitations of 3DGS, their reliance on screen space computations still makes them prone to artifacts, particularly when rendering out-of-distribution camera poses---\ie, viewpoints or parameters significantly different from the training data (\cf \cref{fig:teaser}): 
(1) Even when considering the highest contribution point in 3D~\cite{yu2024gof,radl2024stopthepop} or adjusting transformations per Gaussian~\cite{wu2024_3dgut}, distortions may still occur. 
(2) Zooming in or out beyond the typical training views introduces artifacts, especially when evaluating Gaussians in 3D, as 2D anti-aliasing techniques can no longer be applied~\cite{Yu2023MipSplatting}. 
(3) Despite the use of 3D bounding planes, screen space computations become unstable when Gaussians extend behind the image plane, leading to artifacts at image boundaries~\cite{hahlbohm2024htgs}. 
(4) Simplified per-pixel sorting strategies that focus on high-opacity Gaussians, can result in inaccurate renderings, particularly for viewpoints far from the training distribution~\cite{hahlbohm2024htgs}.

We address these shortcomings in our 3D Gaussian rasterizer, which considers the 3D nature of Gaussians through all steps of the 3DGS rendering pipeline, making the following contributions:
\begin{itemize}
    \item We start by analyzing previous anti-aliasing approaches and introduce an adaptive 3D smoothing filter that accurately dilates 3D Gaussians and removes aliasing artifacts, especially for out-of-distribution views.
     \item We show how bounding of 3D Gaussians can be moved from screen space to view space for stable bounding of Gaussians that reach outside the view frustum, removing disturbing popping artifacts on image boundaries.
     \item  Elevating previous 2D tile-based culling algorithms to 3D by performing frustum-based culling with screen space planes, thereby accelerating rendering and reduce sorting costs for depth-sorted hierarchical rasterization~\cite{radl2024stopthepop}.
\item A detailed analysis of common 3D Gaussian rendering artifacts, and ablation of our employed components.
\end{itemize}
Overall, our method achieves state-of-the-art reconstruction quality, while enabling artifact-free rendering in real-time.
Our training and rendering source code is publicly available at \href{https://github.com/DerThomy/AAA-Gaussians}{https://github.com/DerThomy/AAA-Gaussians}.

%% file: 2_related_work.tex
\section{Related Work}
\label{sec:relwork}

In this section, we cover recent radiance field and 3D Gaussian Splatting methods, with a focus on artifact-free rendering of 3D Gaussian representations.

\subsection{Radiance Fields \& 3D Gaussian Splatting}

Radiance fields have attracted widespread interest in novel view synthesis since the publication of Neural Radiance Fields (NeRF)~\cite{Mildenhall2020NeRF}, which use large coordinate-based MLPs to model view-dependent color and density.
Follow-up work includes improvements in terms of anti-aliasing~\cite{barron2021mipnerf, barron2023zip}, extending NeRFs to unbounded scenes~\cite{barron2022mipnerf360,neff2021donerf}, more efficient encodings~\cite{mueller2022instant}, or fast rendering~\cite{fridovich2022plenoxels, duckworth2023smerf, steiner2024nerfcaching}. 
Despite the aforementioned improvements, most NeRF methods still require multiple costly MLP evaluations for each pixel, manifesting in long training and rendering times.

More recently, 3D Gaussian Splatting (3DGS)~\cite{kerbl3Dgaussians} exploded in popularity, replacing the slower implicit representations of NeRFs with an explicit 3D Gaussian point cloud representation that can be efficiently rasterized through elliptical weighted average (EWA) splatting~\cite{zwicker_ewa_2001}.
The initial sparse point cloud can either be initialized randomly, from Structure-from-Motion~\cite{schoenberger2016sfm}, or from other guidances (\eg, a pretrained NeRF~\cite{niemeyer2024radsplat}).
To achieve good scene coverage, this point cloud has to be densified by adding, cloning, or pruning points based on screen space gradients~\cite{kerbl3Dgaussians}, per-view saliency maps~\cite{mallick2024taming}, depth supervision~\cite{kerbl2024hierarchical}, or by relocating low-opacity Gaussians~\cite{kheradmand2024mcmc}.
Additionally, strategies such as opacity decay~\cite{radl2024stopthepop} and iterative pruning~\cite{fang2024minisplatting, fan2023lightgaussian} are employed to reduce the size of the resulting point cloud.

\subsection{Artifacts in 3D Gaussian Splatting}

The original 3DGS suffers from a number of artifacts (\underline{underlined}), which have been addressed in recent related work.
\underline{Aliasing artifacts} during rasterization can be solved by approximating the integral over each pixel~\cite{liang2024analytic}, using multi-scale 3D Gaussians~\cite{yan2024multi}, or by applying a smoothing filter to both the 3D Gaussian and its projected 2D splat~\cite{Yu2023MipSplatting}.
\underline{Popping artifacts} occur due to the global sort of primitives before rasterization, leading to sudden changes in the blending order.

StopThePop~\cite{radl2024stopthepop} resolves this by computing Gaussian depth values along each view ray and using hierarchical $k$-buffers to improve the sort order.
Other recent work relies on hybrid transparency~\cite{maule2013hybrid}, where only important contributors are sorted, while low-opacity Gaussians are blended in a "tail"~\cite{hahlbohm2024htgs}.
Even perfect per-pixel sorting exhibits \underline{blending artifacts}, due to the non-overlapping assumption of primitives, however, this cannot be solved analytically for Gaussians~\cite{mai2024everexactvolumetricellipsoid} and requires expensive volumetric integration~\cite{condor2025dontsplat, celarek2025does}.
Furthermore, \underline{projection artifacts} occur due to the affine approximation when projecting 3D Gaussians to 2D splats, resulting in disturbing cloud-like artifacts and elongated Gaussians, especially at the image border and for large field of view settings (\eg in virtual reality~\cite{tu2025vrsplat}).
Recent methods solve this by projecting Gaussians onto the unit sphere tangent plane~\cite{huang2024optimal} or using Unscented Transform~\cite{wu2024_3dgut}.
Another way to avoid projection artifacts is to evaluate Gaussians in 3D.
Several methods use ray tracing for this purpose~\cite{moenneloccoz2024gaussianraytracing, mai2024everexactvolumetricellipsoid, byrski2025raysplats}, however, these methods are computationally expensive and require additional acceleration data structures, as well as specialized ray tracing hardware to achieve competitive performance.
To circumvent this, several methods propose to bound the Gaussian on screen, and then evaluate it in 3D by finding its point of maximum contribution along the ray.
However, these methods rely on inaccurate 2D bounds from the approximate affine projection~\cite{yu2024gof,radl2024stopthepop,talegaonkar2025volumetric} or Unscented Transform~\cite{wu2024_3dgut}.
Only recently, \citet{hahlbohm2024htgs} computed accurate 2D bounds through plane fitting to the 3D Gaussian ellipsoid~\cite{botsch2006quadraticsurfaces}.

Despite the large body of work addressing artifact-free 3D Gaussians rendering, many of the aforementioned works imply significant performance penalties, still exhibit errors in extreme configurations or focus only on a single artifact.
To the best of our knowledge, our method is the first to tackle all artifacts in a single, unified framework.

%% file: 3_method.tex
\section{Method}
\label{sec:method}

This section covers necessary preliminary knowledge, and the components of our artifact-free 3D Gaussian renderer.
Our rasterizer utilizes hierarchical sorting~\cite{radl2024stopthepop} and 3D Gaussian evaluation through screen space planes~\cite{hahlbohm2024htgs} to remove popping and projection artifacts.
Sec. \ref{sec:method_antialiasing} introduces our novel adaptive 3D filter that removes aliasing artifacts; Sec. \ref{sec:method_correct_bounding} contains our perspective correct bounding approach that prevents pop-in of Gaussians from outside the view-frustum; Finally, Sec. \ref{sec:method_frustum_culling} proposes our novel frustum-based culling, a crucial component that improves both performance and sort-order through hierarchical culling.

\subsection{Preliminaries}
\paragraph{Gaussian Splatting.}
Following \citet{kerbl3Dgaussians}, we model the scene as a collection of anisotropic 3D Gaussians that serves as an approximation of the scene's geometric structure. Each Gaussian is parameterized by its 3D mean $\bm{\upmu} \in \mathbb{R}^3$, scaling factor $\mathbf{s} \in \mathbb{R}_+^3$, and rotation quaternion $\mathbf{q} \in \mathbb{R}^4$. The probability density function of a Gaussian at a position $\mathbf{x} \in \mathbb{R}^3$ is defined as
\begin{align}
    \mathcal{G}(\mathbf{x}) = \exp\left(-\frac{1}{2} \rho(\mathbf{x})^2\right) \,  \text{, using} \\
    \rho(\mathbf{x}) = \sqrt{(\mathbf{x} - \bm{\upmu})^\top \mathbf{\Sigma}^{-1} (\mathbf{x} - \bm{\upmu})} \, \text{.}
\end{align}
The covariance matrix $\mathbf{\Sigma}$ is given by
\begin{equation}
    \mathbf{\Sigma} = \mathbf{R} \mathbf{S} \mathbf{S}^\top \mathbf{R}^\top \, \text{,}
\end{equation}
where $\mathbf{S}=\text{diag}(\mathbf{s})$ is a diagonal scaling matrix and $\mathbf{R}$ is the rotation matrix obtained from $\mathbf{q}$.
For rendering, the 3D Gaussians are projected onto the 2D image plane.
Since Gaussian distributions are not inherently preserved under nonlinear transformations, \citet{kerbl3Dgaussians} employ a local affine approximation~\cite{zwicker_ewa_2001}.

\paragraph{Evaluation in 3D.}
This approximation introduces a projection error that leads to rendering artifacts, especially at image borders~\cite{huang2024optimal}.
To circumvent these issues, \citet{hahlbohm2024htgs} evaluate Gaussians directly in 3D:
They represent the ray through each pixel $(x, y)$ as the intersection of the two screen space planes $\bm{\uppi}_x = (1, 0, 0, -x)^\top$ and $\bm{\uppi}_y = (0, 1, 0, -y)^\top$, which are then transformed into each Gaussian's normalized space, where the distance to the origin is equal to $\rho(\mathbf{x})$:
\begin{equation}
    \mathbf{T}' = \mathbf{M}_{\text{vp}} \mathbf{P} \mathbf{V} \mathbf{T} \quad \text{using} \quad \mathbf{T} = \begin{pmatrix}
        \mathbf{R} \mathbf{S} & \bm{\upmu} \\
        \mathbf{0}^\top & 1
    \end{pmatrix} \, \text{,}
\end{equation}
where $\mathbf{T}$ is the transformation from Gaussian space to world space, while $\mathbf{M}_{\text{vp}}$, $\mathbf{P}$ and $\mathbf{V}$ correspond to the viewport, projection and view matrices, respectively.
The transformation of the planes is therefore given by
\begin{equation}
    \bm{\uppi}'_{x/y} = \left(\mathbf{T'}^{-1}\right)^{-\top} \bm{\uppi}_{x/y} = \mathbf{T}'^\top \bm{\uppi}_{x/y} \, \text{.}
\end{equation}

By computing the distance of the intersection line formed by the two planes to the origin, the Gaussian’s contribution to the pixel ray is determined. 
In this formulation, the integral is no longer evaluated to determine the Gaussian contribution; instead, only the maximum contribution along the ray is considered.

This formulation avoids the use of the inverse covariance $\mathbf{\Sigma}^{-1}$, which can become numerically instable for degenerate Gaussians (any $\mathbf{s}_i \approx 0$).

\begin{figure}[t]
    \centering
    \begin{subfigure}[]{0.495\linewidth}\includegraphics[width=\linewidth]{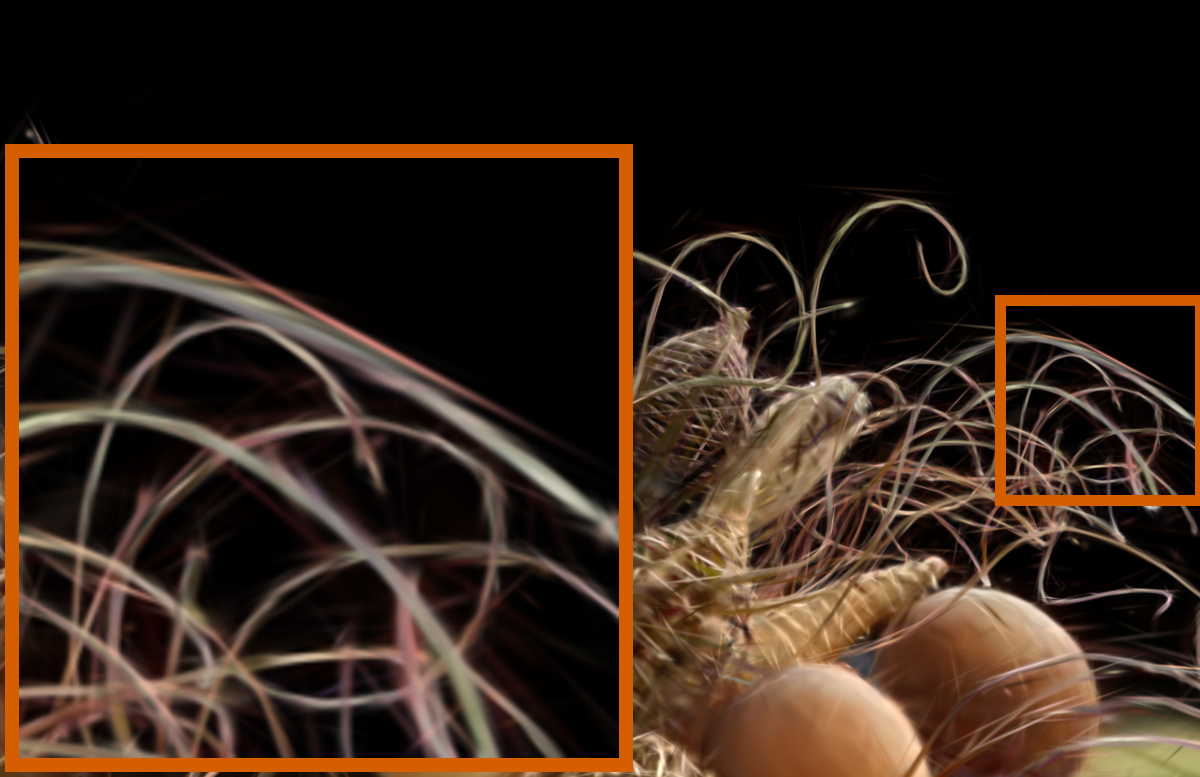}\end{subfigure} \hfill
    \begin{subfigure}[]{0.495\linewidth}\includegraphics[width=\linewidth]{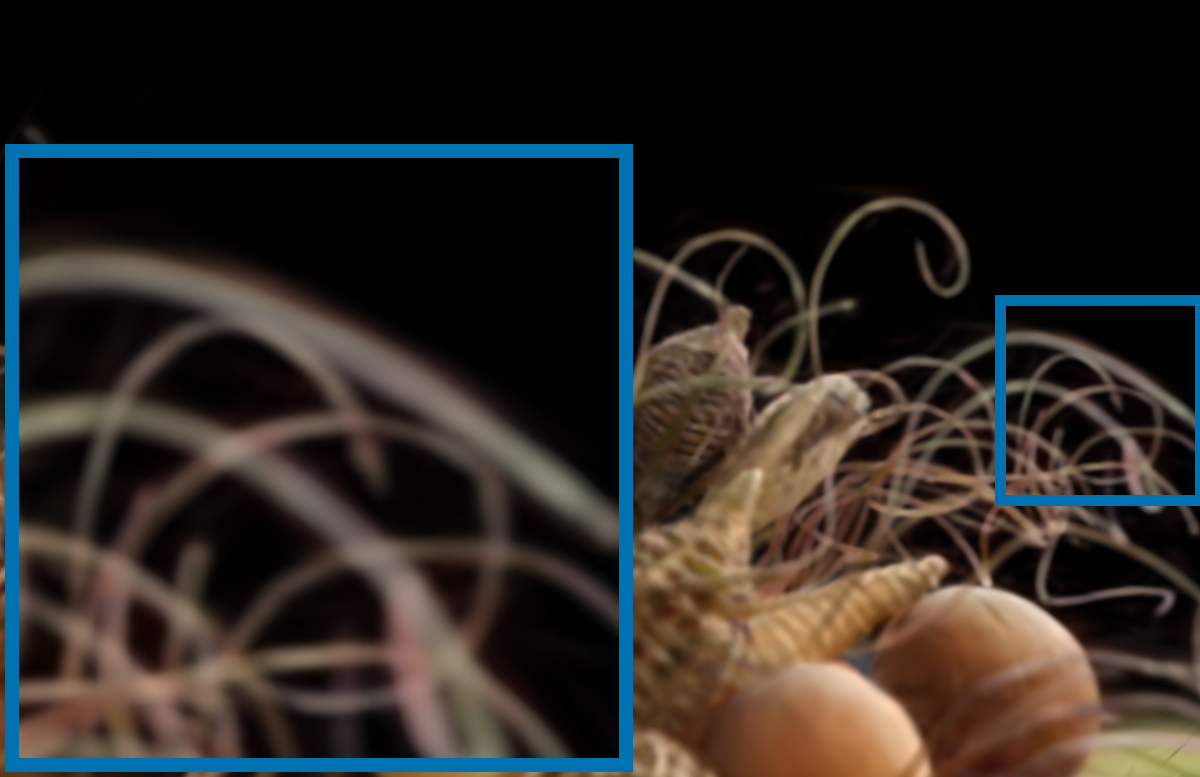}\end{subfigure} %
    \par\vspace{1.5pt}
    \begin{subfigure}[]{0.495\linewidth}\includegraphics[width=\linewidth]{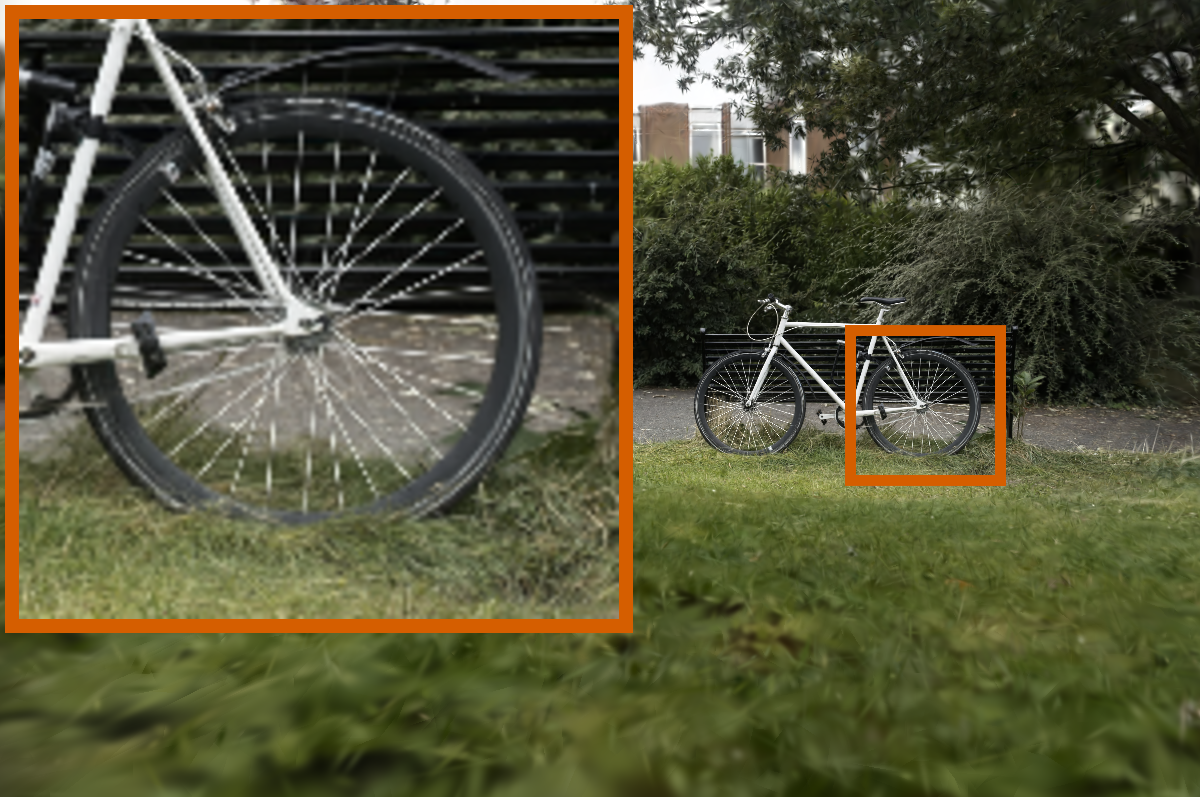}\end{subfigure} \hfill
    \begin{subfigure}[]{0.495\linewidth}\includegraphics[width=\linewidth]{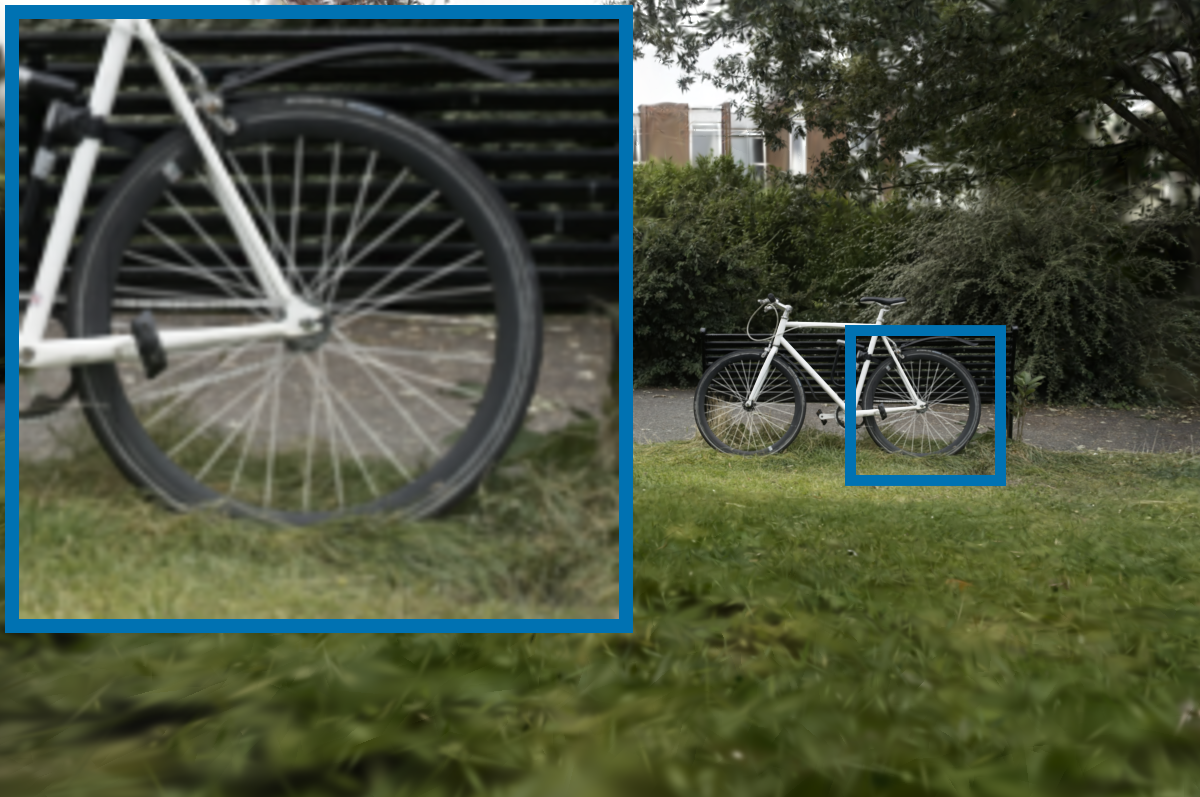}\end{subfigure}
    \caption{(Left) Aliasing artifacts manifest when camera positions deviate significantly from training distances: (1) Gaussians become too thin due to over-sampling when moving close. (2) Gaussians become too small due to under-sampling when moving farther away.
    (Right) By dynamically adjusting to varying view conditions, our adaptive 3D filter effectively removes these artifacts, preserving fine details and ensuring consistent image quality.}
    \label{fig:aliasing_filter}
\end{figure}

\paragraph{Anti Aliasing.}
Aliasing artifacts arise when the sampling rate during rendering deviates from the one used during training, \eg changes in image resolution, focal length, or viewing distance (\cf \cref{fig:aliasing_filter}).
To address this, \citet{Yu2023MipSplatting} combine a 3D smoothing filter with a 2D screen space Mip filter.
The 3D smoothing filter prevents high-frequency sampling artifacts by applying a low-pass Gaussian filter to each Gaussian, where the filter size is determined by the maximum sampling frequency observed during training across all $N_c$ training cameras, given by
\begin{equation}
    \hat{v}_{\text{train}} = \max\left( \left\{ \hat{v}_n \right\}_{n=1}^{N_c} \right), \quad \text{with} \quad \hat{v} = \frac{f}{d} \, \text{,}
\end{equation}
where $f$ denotes the camera's focal length in pixels and $d$ represents the $z$-component of the Gaussian's mean $\bm{\upmu}$ in view space.
The smoothing filter is defined by covariance $\hat{\mathbf{\Sigma}} = \mathbf{\Sigma} + \nicefrac{k}{\hat{v}^2} \  \mathbf{I}$, where $k$ controls the size of the filter. The smoothed Gaussian is then given by
\begin{equation}
    \label{eq:gauss_dil}
    \hat{\mathcal{G}}(\mathbf{x}) = \sqrt{\frac{|\mathbf{\Sigma}|}{|\mathbf{\hat{\Sigma}}|}} \ \exp\left(-\frac{1}{2} (\mathbf{x} - \bm{\upmu})^\top \mathbf{\hat{\Sigma}}^{-1} (\mathbf{x} - \bm{\upmu})\right) \, \text{.}
\end{equation}

To prevent low-frequency sampling artifacts---\eg when increasing the camera distance---they replace the screen space dilation filter of \citet{kerbl3Dgaussians} with a 2D Mip filter, which approximates a 2D box filter in image space.

\paragraph{Tile-Based Rendering.}
To optimize rendering efficiency, 3DGS partitions the image into tiles.
Each Gaussian is evaluated for overlap with these tiles and assigned accordingly, ensuring that each tile processes only the Gaussians that intersect with it during rendering.
To determine overlap, \citet{kerbl3Dgaussians} compute a screen space bounding box based on the eigenvalues of the 2D covariance matrix.

\paragraph{Blending Sort Order.}
\citet{kerbl3Dgaussians} approximate the blending order based on the global depth of each Gaussian's view space mean.
This strategy eliminates the need for a computationally expensive per-pixel sorting of Gaussians, but introduces popping artifacts when the camera rotates. To address this issue, \citet{radl2024stopthepop} introduce an efficient hierarchical rasterization approach that approximates a pixel-perfect sorting, significantly reducing visual artifacts while maintaining computational efficiency by interleaving hierarchical sorting with repeated culling.

\subsection{3D Gaussian Anti-Aliasing}
\label{sec:method_antialiasing}

\begin{figure}[t]
    \centering
    
    \begin{subfigure}[b]{0.495\linewidth}
        \includegraphics[width=\linewidth]{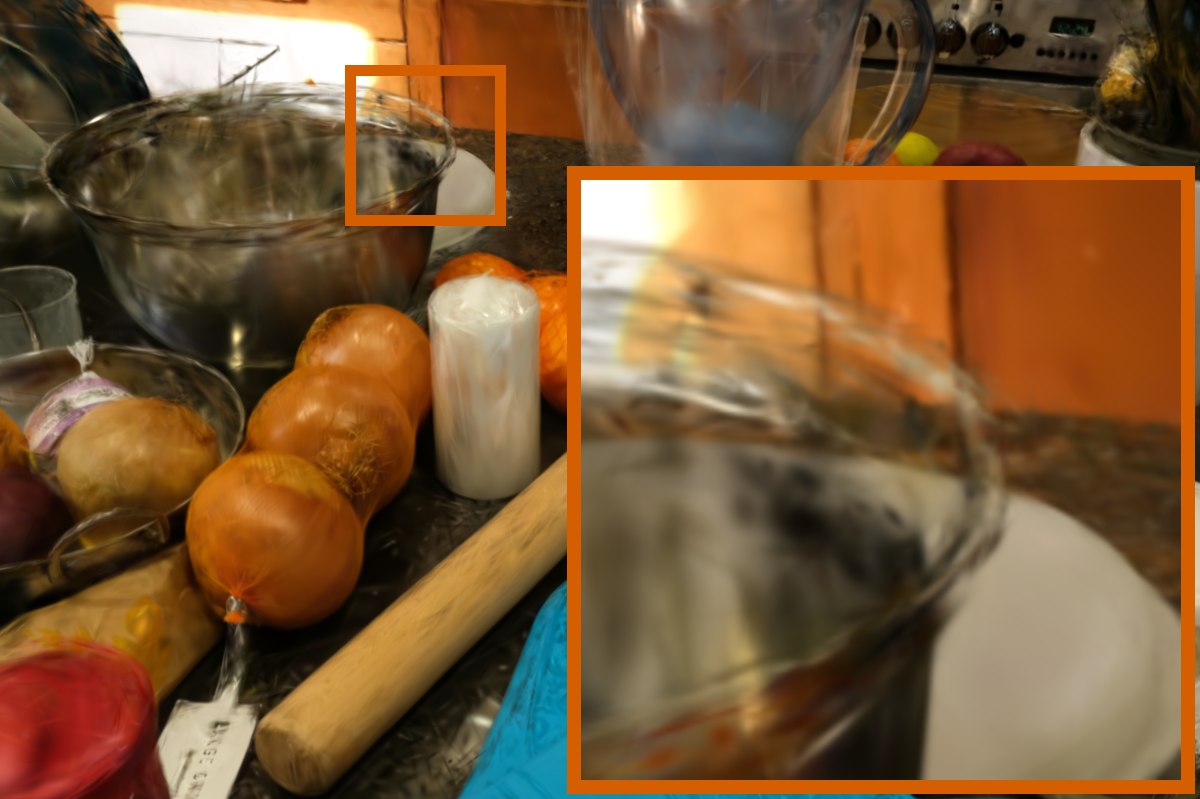}
    \end{subfigure}
    \hfill
    \begin{subfigure}[b]{0.495\linewidth}
        \includegraphics[width=\linewidth]{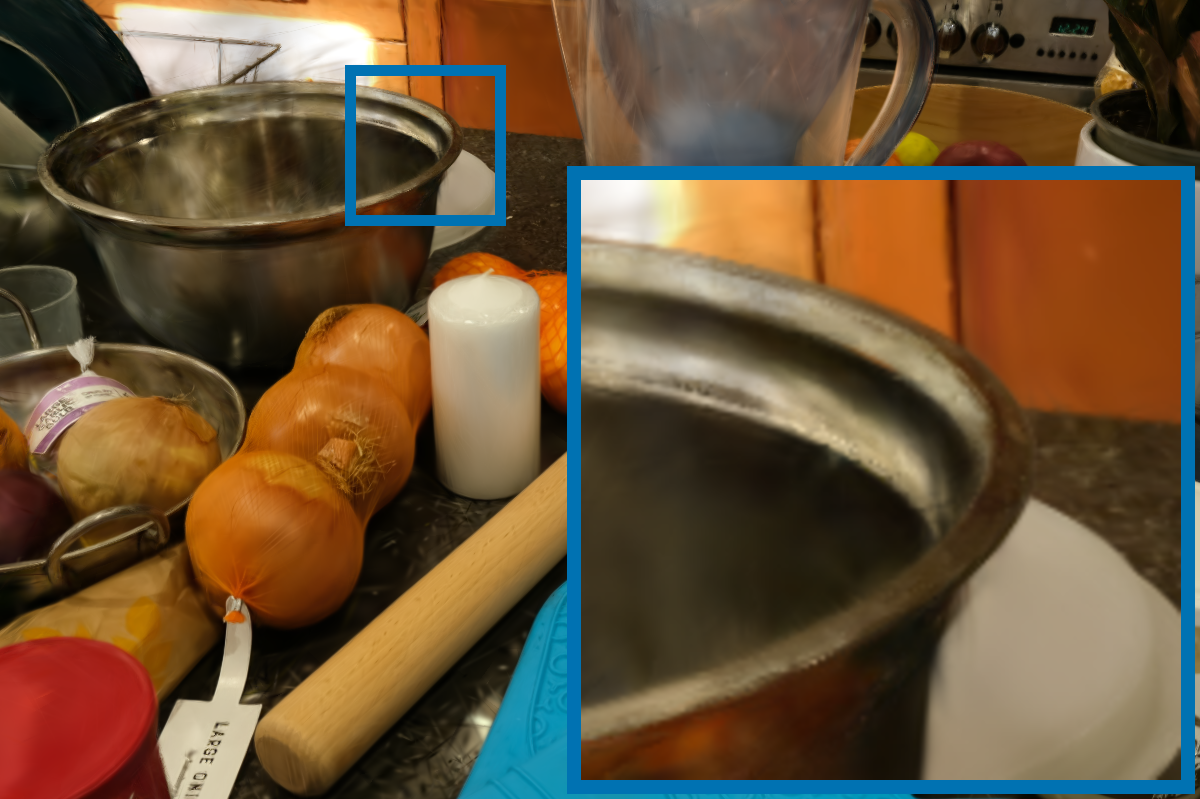}
    \end{subfigure}

    \vspace{-0.5em}
    
    \begin{tikzpicture}
        \draw[dashed] (0,0) -- (\linewidth,0);
    \end{tikzpicture}

    \begin{subfigure}[b]{0.495\linewidth}
        \includegraphics[width=\linewidth]{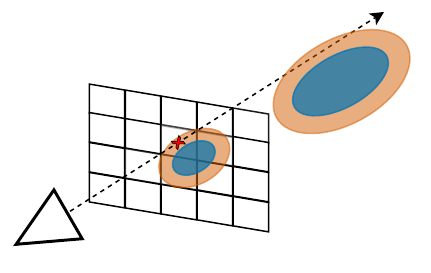}
    \end{subfigure}
    \hfill
    \begin{subfigure}[b]{0.495\linewidth}
        \includegraphics[width=\linewidth]{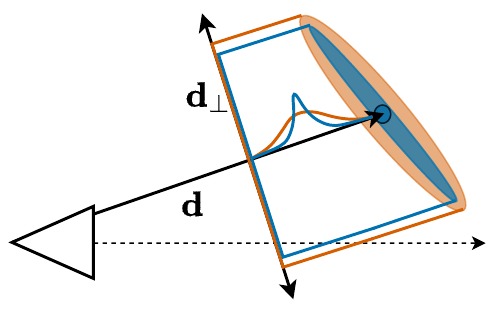}
    \end{subfigure}
    
    \caption{Improvement due to our adaptive 3D dilation filter: Whereas previous methods adjust amplitudes based solely on the change of volume of the 3D Gaussian, often leading to excessive transparency (top left), our method adapts based on the area perpendicular to the viewing ray.}
    \label{fig:proper_dilation}
\end{figure}
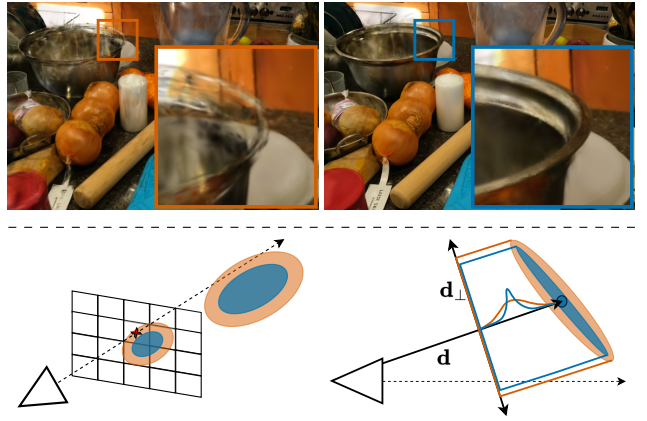

Aliasing is one of the most pressing challenges in 3D Gaussian evaluation, particularly when rendering scenes at varying distances. While the 3D smoothing filter introduced by \citet{Yu2023MipSplatting} is inherently compatible with 3D evaluation, the 2D screen space Mip filter cannot be directly applied. As a result, recent approaches that rely on 3D evaluation omit the screen space filter and depend solely on the 3D smoothing filter~\cite{hahlbohm2024htgs, yu2024gof}. While this effectively mitigates artifacts when moving the camera closer, it remains susceptible to aliasing when increasing the viewing distance, as fine details are not adequately filtered, leading to flickering and loss of visual stability.

To address this limitation, we replace the 2D screen space Mip filter by a full 3D filtering approach that seamlessly integrates with 3D evaluation methods, effectively  preventing low-frequency aliasing. A na\"ive  approach would be to recompute the 3D smoothing filter for each rendering view, but this proves insufficient, as it causes Gaussians to become overly transparent (\cf \cref{fig:proper_dilation}). This issue arises because the amplitude in \cref{eq:gauss_dil} decreases according to the change in volume, while Gaussians are evaluated only at their point of maximum contribution along a ray $\mathbf{d}$, rather than being fully integrated along the ray. Consequently, incorporating the scaling change along the ray overestimates amplitude scaling for highly anisotropic Gaussians, which leads to an excessively small normalization factor
\begin{equation}
    \label{eq:detprod}
    \sqrt{\frac{|\mathbf{\Sigma}|}{|\mathbf{\hat{\Sigma}}|}} = \sqrt{\frac{\prod_{i=1}^3 \mathbf{s}^2_i}{\prod_{i=1}^3 \left(\mathbf{s}^2_i + \frac{k}{\hat{v}^2} \right)}} \, \text{,}
\end{equation}
in \cref{eq:gauss_dil}.
Notably, this reduction occurs even when the scale change perpendicular to $\mathbf{d}$ is minimal, highlighting the need for a more robust filtering approach. Therefore, we reformulate the normalization to only factor in the change of scale perpendicular to $\mathbf{d}$ (\cf \cref{fig:proper_dilation}), that is
\begin{equation}
    \hat{\mathcal{G}}_\perp(\mathbf{x}) = \sqrt{\frac{|\mathbf{\Sigma}_\perp|}{|\mathbf{\hat{\Sigma}}_\perp|}} \ \exp\left(-\frac{1}{2} (\mathbf{x} - \bm{\upmu})^\top \mathbf{\hat{\Sigma}}^{-1} (\mathbf{x} - \bm{\upmu})\right) \, \text{,}
\end{equation}
where $\mathbf{d}$ is the normalized vector between $\bm{\upmu}$ and the camera origin $\mathbf{o}$, and $\mathbf{\Sigma}_\perp$ denotes the $2 \times 2$ covariance matrix projected onto the subspace orthogonal to $\mathbf{d}$. It can be shown (\cf the supplementary material for a derivation) that the perpendicular scaling factor is given by
\begin{equation}
    \label{eq:perpscale}
    \sqrt{\frac{|\mathbf{\Sigma}_\perp|}{|\mathbf{\hat{\Sigma}}_\perp|}} = \sqrt{\frac{|\mathbf{\Sigma}| \ \mathbf{d}^\top \mathbf{\Sigma}^{-1} \mathbf{d}}{|\mathbf{\hat{\Sigma}}| \ \mathbf{d}^\top \mathbf{\hat{\Sigma}}^{-1} \mathbf{d}}} \, \text{.}
\end{equation}
Since the inverse covariance matrix is given by $\mathbf{\Sigma}^{-1} = \mathbf{R} \mathbf{S}^{-2} \mathbf{R}^\top$, we can express the directional quadratic form as
\begin{equation}
    \mathbf{d}^\top \mathbf{\Sigma}^{-1} \mathbf{d} = \mathbf{d}^\top \mathbf{R} \ \mathbf{S}^{-2} \mathbf{R}^\top \mathbf{d} = \sum_{i = 1}^3 \frac{\mathbf{d}'^2_i}{\mathbf{s}^2_i} \, \text{,}
\end{equation}
where $\mathbf{d}' = \mathbf{R}^\top \mathbf{d}$. Using the determinant factorization from \cref{eq:detprod}, we can simplify \cref{eq:perpscale} to
\begin{equation}
    \sqrt{\frac{\mathbf{d}'^2_1 \ \mathbf{s}^2_2 \ \mathbf{s}^2_3 + \mathbf{d}'^2_2 \ \mathbf{s}^2_1 \ \mathbf{s}^2_3 + \mathbf{d}'^2_3 \ \mathbf{s}^2_1 \ \mathbf{s}^2_2}{\mathbf{d}'^2_1 \ \mathbf{\hat{s}}_2 \ \mathbf{\hat{s}}_3 + \mathbf{d}'^2_2 \ \mathbf{\hat{s}}_1 \ \mathbf{\hat{s}}_3 + \mathbf{d}'^2_3 \ \mathbf{\hat{s}}_1 \ \mathbf{\hat{s}}_2}} \, \text{,}
\end{equation}
where the updated scaling factors are given by $\mathbf{\hat{s}}_i = \mathbf{s}^2_i + \nicefrac{k}{\hat{v}^2}$. This formulation enables efficient computation while avoiding explicit matrix inversion, ensuring numerical stability. 
To address artifacts when moving the camera close to a Gaussian, we integrate the 3D smoothing filter of \citet{Yu2023MipSplatting} with our proposed 3D kernel:
We store the maximum sampling frequency observed across all training cameras $\hat{v}_\text{train}$. During rendering, the effective sampling frequency is defined as
\begin{equation}
    \hat{v}' = \min(\hat{v}_\text{train}, \hat{v}) \, \text{,}
\end{equation}
where $\hat{v}'$ corresponds to the sampling frequency used for our filter. This formulation ensures that Gaussians do not shrink excessively when the camera moves closer, while still providing effective anti-aliasing when the camera moves farther away.

\subsection{Perspective Correct Bounding}
\label{sec:method_correct_bounding}

\begin{figure}
    \captionsetup{type=figure}
    \begin{subfigure}{0.22\textwidth}
    \centering
    \includegraphics[width=0.9\textwidth]{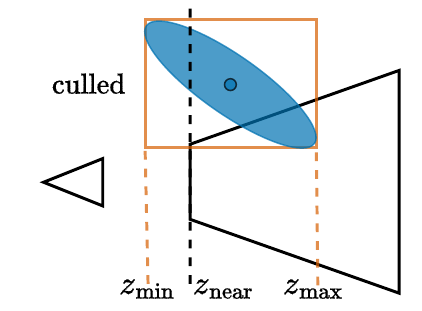}
\end{subfigure}
    \begin{subfigure}{0.22\textwidth}
    \centering    
     \raisebox{1.65mm}{\includegraphics[width=0.8\textwidth]{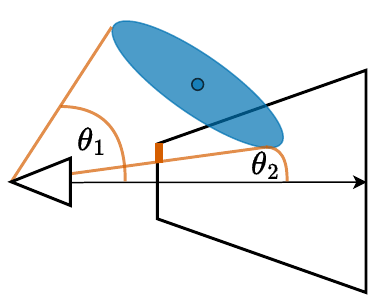}}
    \end{subfigure}

    \captionof{figure}{(Left) \citet{hahlbohm2024htgs} compute the screen bounds of a Gaussian by fitting planes in screen space.
    However, they discard Gaussians whose $z$-bounds ($z_{\min,\max}$) are outside the near/far-planes, which can lead to popping.
    (Right) We instead compute view space angles $\theta_{1,2}$, leading to a more robust computation and bounding.}
    \label{fig:bounding_diagrams}
\end{figure}

Efficient evaluation of 3D Gaussians in a software rasterizer requires accurate bounding in screen space to avoid unnecessary evaluations.
Following \citet{botsch2006quadraticsurfaces}, \citet{hahlbohm2024htgs} perform exact plane fitting to the ellipsoid, defined by the 3D Gaussian's level set at $\tau_\rho$ in projective space, which fails when a Gaussian's extent reaches behind the image plane.
To mitigate this, they discard these Gaussians, leading to noticeable popping (\cf Fig.~\ref{fig:bounding_diagrams}).
Instead, we perform the plane fitting in view space with planes $\bm{\uppi}_\theta=(\cos(\theta),0,-\sin(\theta),0)^\top$, and $\bm{\uppi}_\phi=(0,\cos(\phi),-\sin(\phi),0)^\top$, and solving for $\theta$ and $\phi$:
\begin{align}
    \theta_{1,2}=\tan^{-1} \left( \frac{s_{1,3} \pm \sqrt{s_{1,3}^2 - s_{1,1}s_{3,3}}}{s_{3,3}} \right), \\
    \phi_{1,2}=\tan^{-1} \left( \frac{s_{2,3} \pm \sqrt{s_{2,3}^2 - s_{2,2}s_{3,3}}}{s_{3,3}} \right),
\end{align}
with $s_{i,j}=\langle \mathbf{t}, \mathbf{T}_{(\text{view},i)} \odot \mathbf{T}_{(\text{view},j)}\rangle$, $\bm{t}=(\tau_\rho,\tau_\rho,\tau_\rho,-1)^\top$, and $\mathbf{T}_{(\text{view},i)}$ denoting the $i$-th row of transformation matrix from Gaussian space to view space $\mathbf{T}_\text{view} = \mathbf{V}\mathbf{T}$ (\cf the supplementary material for a derivation).
We additionally compute the angles $\theta_\mu,\phi_\mu$ of the view space Gaussian mean in $x/y$-direction. This is followed by a rotation step, where we ensure that
\begin{align}
    (\theta_\mu - \pi) < \theta_1 < \theta_\mu < \theta_2 < (\theta_\mu + \pi),
\end{align}
and a bounding step to the range $[-(\frac{\pi}{2}-\epsilon), \frac{\pi}{2}-\epsilon]$ (with a small $\epsilon\in\mathbb{R}_+$) to translate those bounds to the screen 
\begin{align}
    \theta_1 = \max \left(-\frac{\pi}{2} + \epsilon, \theta_1 \right), \theta_2 = \min \left( \frac{\pi}{2} - \epsilon, \theta_2 \right).
\end{align}
If the camera center is inside the Gaussian's ellipsoid, no valid solution exists, in which case we discard this Gaussian.
Additionally, if the ellipsoid intersects the $x$-axis, it cannot be bounded in the screen space $y$-axis, and vice-versa.
In these cases, the term inside the square root becomes negative, and we conservatively set the bounds to the entire screen for the affected axis.
In the next step, tile-based culling removes all tiles that receive negligible contribution from the Gaussian or where the Gaussian's depth remains closer than the near plane across the entire tile.

\begin{figure}
    \includegraphics[width=0.23\textwidth]{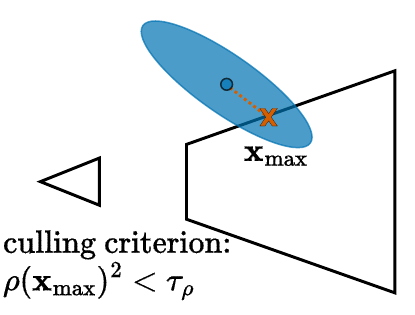}
    \includegraphics[width=0.23\textwidth]{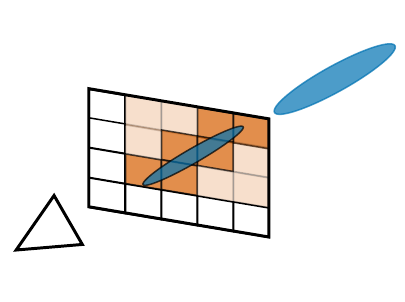}
    \captionof{figure}{Our frustum culling algorithm finds the maximum contribution point inside a 3D frustum, via projection of the origin onto the transformed planes and edges in Gaussian space.
    Comparing $\rho(\mathbf{x})$ at this maximum point against the threshold $\tau_\rho$, we can cull away a whole Gaussian against the view frustum (Left), as well as individual tiles for a single Gaussian (Right).}
    \label{fig:frustum_culling}
\end{figure}

\subsection{Frustum-Based Culling} 
\label{sec:method_frustum_culling}

Accurate screen space bounding with axis-aligned bounding boxes (AABBs) of the ellipsoids helps to reduce the per-pixel workload, however, it delivers bad bounds for highly non-axis-aligned ellipsoids.
Additionally, Gaussians could receive valid screen bounds, but never contribute to any pixel.
To mitigate this, \citet{radl2024stopthepop} perform per-tile culling of the projected 2D ellipse, which drastically reduces the number of Gaussian/tile combinations, and also prevents them from performing unnecessary sorting operations in their hierarchical sort.
We translate this tile-based culling approach to 3D by constructing a per-tile frustum $\mathcal{F}$ from 4 planes $\bm{\uppi}_{x_{1,2}}=(1, 0, 0, -x_{(\min,\max)}), \bm{\uppi}_{y_{1,2}}=(0,1,0,-y_{(\min,\max)})$, where $x_{(\min,\max)}, y_{(\min,\max)}$ define the tile boundaries in pixel coordinates.
We then compute the point of maximum contribution of the Gaussian inside this 3D frustum and discard tiles where $\rho(\mathbf{x})^2$ is above the threshold $\tau_\rho$, \ie
\begin{align}
    \min_{\mathbf{x}\in\mathcal{F}} \rho(\mathbf{x})^2 < \tau_\rho.
\end{align}

In the trivial case where the Gaussian's mean is already inside this frustum, it is consequently the point of maximum contribution.
Otherwise, this point has to lie on the planes and edges of the frustum.
We find the maximum contribution point on the planes by transforming them into the normalized Gaussian space, and finding the point closest to the origin.
A na\"ive solution is to do this for each plane and edge, and ensuring that the projected point lies within the bounds of the frustum and in front of the camera.
Instead, we only project onto the $x/y$-planes (and their corresponding edges) that are closest to the Gaussian's mean in screen space, limiting the evaluations to 2 planes and 3 edges (instead of 4 planes and 4 edges for the na\"ive approach).
This can be implemented efficiently, which is critical as this routine has to be executed for many tiles per Gaussian.

Additionally, we cull Gaussians during pre-processing against the entire view frustum to discard all non-contributing Gaussians (\cf Fig.~\ref{fig:frustum_culling}).
In contrast to our exact frustum culling, other methods discard Gaussians only based on their mean (\eg, if it lies behind the image plane~\cite{kerbl3Dgaussians}) or based on the screen space $z$-bounds~\cite{hahlbohm2024htgs}, which results in incorrect culling of contributing Gaussians and popping artifacts at the image borders.

%% file: 4_evaluation.tex
\section{Evaluation}
\label{sec:eval}

Following  prior work, we evaluate our method on 13 outdoor and indoor scenes from three different datasets: Mip-NeRF~360~\cite{barron2022mipnerf360}, Tanks~\&~Temples~\cite{Knapitsch2017tanks}, and Deep Blending~\cite{hedman2018deepblending}.

\paragraph{Implementation Details.}
We use the pre-downscaled images of Mip-NeRF~360~\cite{barron2022mipnerf360} for training and evaluation, following the setup of 3DGS~\cite{kerbl3Dgaussians}.
For densification, we adopt Markov Chain Monte Carlo (MCMC)~\cite{kheradmand2024mcmc}  with identical parameter settings.
Our approach builds on the hierarchical rasterizer of \citet{radl2024stopthepop}, retaining their per-ray sorting and queuing strategies while replacing the bounding, culling, depth evaluation, contribution estimation, and anti-aliasing with our 3D-aware implementations.
All compared methods optimize to the same number of primitives, with the exception of Hybrid Transparency~\cite{hahlbohm2024htgs}, where we re-evaluate image metrics from their original results, as their training code was unavailable at the time of writing. Following previous works~\cite{zwicker_ewa_2001, kerbl3Dgaussians, Yu2023MipSplatting} we use a kernel size of $k = 0.3$ for our adaptive 3D filter.

\subsection{Image Metrics}
We compare our method against 3DGS~\cite{kerbl3Dgaussians}, MCMC~\cite{kheradmand2024mcmc}, and Taming 3DGS~\cite{mallick2024taming}, which all use different densification approaches.
Additionally, we compare against StopThePop~\cite{radl2024stopthepop} and Mip-Splatting~\cite{Yu2023MipSplatting}, both of which build on the original 3DGS densification but specifically target artifact removal.
For Hybrid Transparency~\cite{hahlbohm2024htgs}, we re-evaluate image metrics on their provided results for the Mip-NeRF360 dataset.
Our evaluation considers PSNR, SSIM, and LPIPS \cite{zhang2018unreasonable}, with LPIPS computed on unnormalized images to maintain consistency with prior work.

\input{image_metrics}

\paragraph{Standard Datasets.}
We begin  by evaluating standard datasets, which represent in-distribution camera and view parameters. 
As shown in \cref{tab:evaluation_full}, our method outperforms others in nearly all metrics and matches MCMC in overall quality---while suffering from none of the artifacts.
Notably, our approach prevents the optimizer from "cheating per-view" inconsistencies via popping which explains the slightly lower PSNR  in Tanks~\&~Temples.
In this dataset, MCMC relies on distorted Gaussians and popping  to compensate for large exposure changes, effectively "faking" full-screen adjustments. However, as shown in \cref{fig:collage},  this strategy breaks down for out-of-distribution viewing configurations.

\paragraph{Ablation.}
We analyze the impact of our components in \cref{tab:ablation_components}.
While standard image metrics remain largely unaffected for in-distribution test views, disabling individual components leads to distinct artifacts.
Our anti-aliasing ensures stability across resolution changes and distance to the observed scene content (\cref{tab:ablation_multires}).
Removing hierarchical per-pixel sorting introduces popping artifacts and allows the method to cheat with a view-inconsistent representation (again see \citet{radl2024stopthepop}).
Finally, disabling 3D evaluation results in projection artifacts, particularly noticeable when increasing the field of view (\cref{fig:collage}).
Frustum-based culling has no impact on reconstruction quality, but it plays a significant role in improving performance (\cf Tab. \ref{tab:performance_ablation}).

\input{tab_main_comp_ablation}

\subsection{View-Consistent Rendering}

Densification primarily impacts image metrics on in-distribution test views, as inconsistencies are implicitly learned during training. However, for out-of-distribution views---\eg with a larger field of view (FOV) or changed resolution---these inconsistencies become more apparent in evaluation metrics.

\begin{figure}
    \centering
    \includegraphics[width=0.99\linewidth]{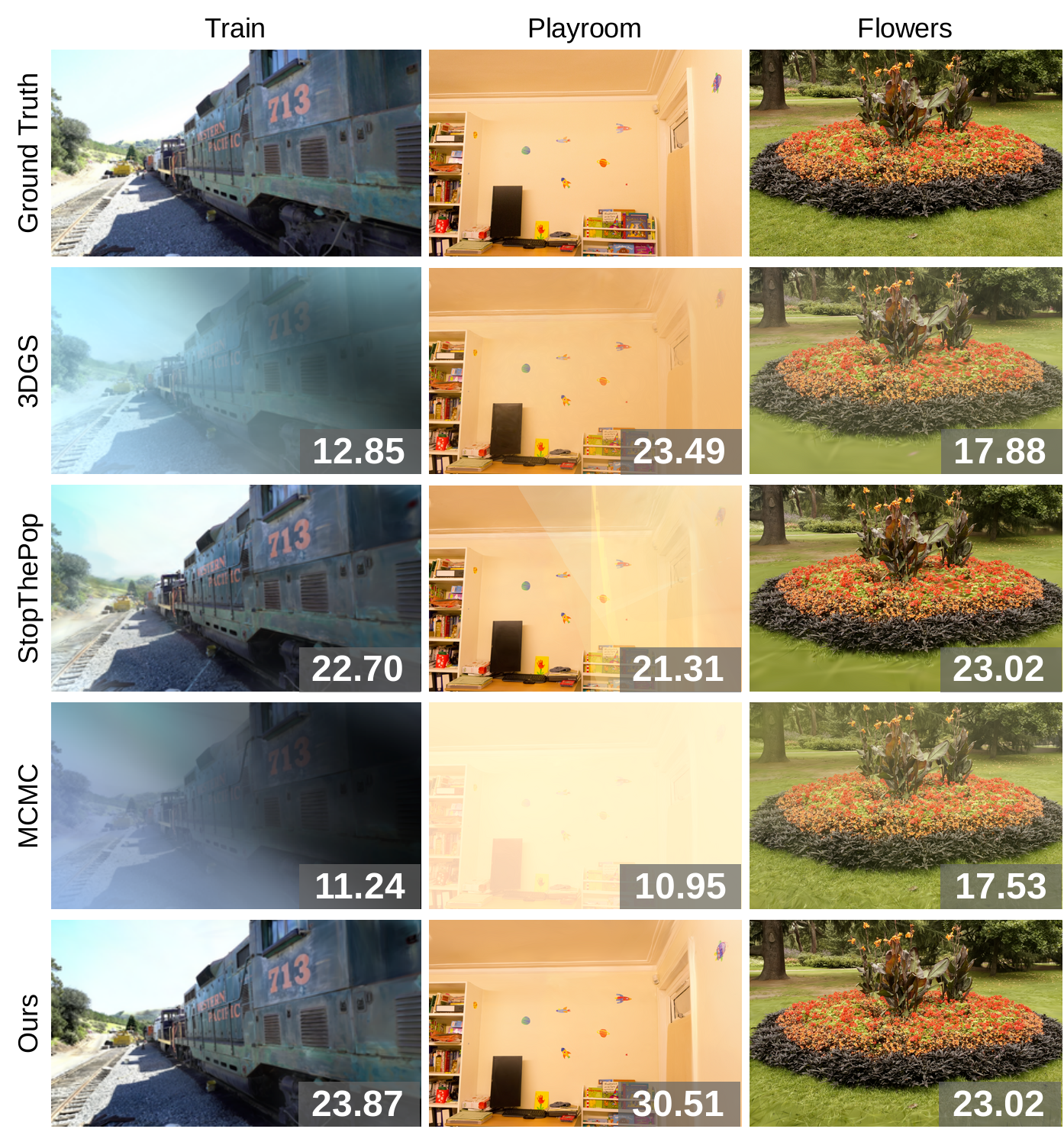}
    \caption{Example results with PSNR when rendering with a larger FOV, comprising out-of-distribution view settings. Clearly, other methods suffer from severe distortion artifacts.}
    \label{fig:collage}
\end{figure}

\paragraph{Larger Field of View.}
We assess FOV robustness by artificially increasing the FOV and resolution of test views while extracting a pixel-perfect cutout from the original image for ground truth comparison. Our setup follows \citet{huang2024optimal} but decreases focal length by $3{\times}$ and increases resolution by $3{\times}$. As shown in \cref{tab:large_fov}, our method remains unaffected by these changes, whereas all other approaches suffer significant quality degradation due to distortion artifacts. A visual comparison is provided in \cref{fig:collage}.

\input{large_fov_tab_singlerowattempt}

\paragraph{Changing Resolution.}
To assess our method’s anti-aliasing capability, we perform a multi-resolution evaluation using the original training resolution ($1{\times}$), half resolution ($\frac{1}{2}{\times}$), and double resolution ($2{\times}$). In \cref{tab:ablation_multires}, we compare results for two Mip-NeRF~360 scenes against the provided pre-downscaled (or original size) images. While all methods perform similarly at $1{\times}$ resolution, our anti-aliasing 3D filter preserves quality across lower and higher resolutions (\cf supplementary material for visual comparisons).

\begin{table}[]
    \centering
    \setlength{\tabcolsep}{3pt}
\resizebox{.98\linewidth}{!}{
\begin{tabular}{clccccccc}
\toprule
\multirow{3}{*}{Res.} & & \multicolumn{3}{c}{Bonsai} && \multicolumn{3}{c}{Bicycle} \\
\cmidrule(lr){3-5} \cmidrule(lr){7-9}
  & & PSNR\textsuperscript{$\uparrow$} & SSIM\textsuperscript{$\uparrow$} & LPIPS\textsuperscript{$\downarrow$} && PSNR\textsuperscript{$\uparrow$} & SSIM\textsuperscript{$\uparrow$} & LPIPS\textsuperscript{$\downarrow$} \\
\midrule
\multirow{3}{*}{\rotatebox{00}{$\frac{1}{2}{\times}$}} & MCMC        & \cellcolor{tab_color!00} 28.98 & \cellcolor{tab_color!00} 0.941 & \cellcolor{tab_color!00} 0.099 && \cellcolor{tab_color!00} 21.15 & \cellcolor{tab_color!00} 0.774 & \cellcolor{tab_color!00} 0.171 \\
                                                     & Ours        & \cellcolor{tab_color!30} 32.12 & \cellcolor{tab_color!30} 0.953 & \cellcolor{tab_color!30} 0.095 && \cellcolor{tab_color!30} 26.73 & \cellcolor{tab_color!30} 0.854 & \cellcolor{tab_color!30} 0.123 \\
                                                     & Ours w/o AA & \cellcolor{tab_color!00} 28.54 & \cellcolor{tab_color!00} 0.939 & \cellcolor{tab_color!00} 0.097 && \cellcolor{tab_color!00} 20.99 & \cellcolor{tab_color!00} 0.771 & \cellcolor{tab_color!00} 0.174 \\
\midrule
\multirow{3}{*}{\rotatebox{00}{$1{\times}$}} & MCMC        & \cellcolor{tab_color!30} 32.65 & \cellcolor{tab_color!30} 0.948 & \cellcolor{tab_color!00} 0.191 && \cellcolor{tab_color!00} 25.69 & \cellcolor{tab_color!00} 0.799 & \cellcolor{tab_color!00} 0.168 \\
                                           & Ours        & \cellcolor{tab_color!00} 32.32 & \cellcolor{tab_color!30} 0.948 & \cellcolor{tab_color!30} 0.189 && \cellcolor{tab_color!30} 25.74 & \cellcolor{tab_color!30} 0.801 & \cellcolor{tab_color!00} 0.171 \\
                                           & Ours w/o AA & \cellcolor{tab_color!00} 32.13 & \cellcolor{tab_color!00} 0.947 & \cellcolor{tab_color!30} 0.189 && \cellcolor{tab_color!00} 25.65 & \cellcolor{tab_color!30} 0.801 & \cellcolor{tab_color!30} 0.165 \\
\midrule
\multirow{3}{*}{\rotatebox{00}{$2{\times}$}} & MCMC        & \cellcolor{tab_color!00} 31.49 & \cellcolor{tab_color!00} 0.936 & \cellcolor{tab_color!00} 0.279 && \cellcolor{tab_color!00} 22.13 & \cellcolor{tab_color!00} 0.689 & \cellcolor{tab_color!00} 0.288 \\
                                           & Ours        & \cellcolor{tab_color!30} 32.09 & \cellcolor{tab_color!30} 0.940 & \cellcolor{tab_color!30} 0.276 && \cellcolor{tab_color!30} 24.52 & \cellcolor{tab_color!30} 0.728 & \cellcolor{tab_color!30} 0.264 \\
                                           & Ours w/o AA & \cellcolor{tab_color!00} 30.99 & \cellcolor{tab_color!00} 0.935 & \cellcolor{tab_color!00} 0.278 && \cellcolor{tab_color!00} 21.90 & \cellcolor{tab_color!00} 0.687 & \cellcolor{tab_color!00} 0.288 \\
\bottomrule
\end{tabular}
}
    \caption{Multi-resolution evaluation ablation of MCMC and our method, with and without anti-aliasing. Ours gives clearly better results when changing the resolution during test time, highlighting the benefits of the 3D anti-aliasing filter.}
    \label{tab:ablation_multires}
\end{table}

\paragraph{Close to Scene Camera Location.}
Popping artifacts caused by incorrect culling when moving close to scene content---which our method resolves---are difficult to evaluate quantitatively, as ground truth data cannot be generated from existing evaluation views.
An example for this kind of popping is shown in \cref{fig:teaser}.
For more examples, please see the supplemental video.

\subsection{Performance Timings}

We evaluate the runtime performance of our components and MCMC in \cref{tab:performance_ablation}, using an NVIDIA RTX 4090 with timings averaged over an interpolated camera path across all available camera poses.
Our method is only slightly slower than MCMC (with standard 3DGS bounding) and even outperforms it on the Tanks and Temples dataset.
When removing culling, performance drops significantly, as the hierarchical sort heavily relies on culling to reduce its sorting overhead.
Disabling hierarchical sorting improves speed beyond MCMC but introduces noticeable popping artifacts. 
Lastly, disabling our 3D evaluation results in similar or worse performance, indicating that our 3D evaluation is as fast or even faster than 2D splat variants.

\begin{table}[]
    \centering
    \small
\begin{tabular}{lrrrr}
\toprule
 & \multirow{2}{*}{\shortstack{M360\\Indoor}} & \multirow{2}{*}{\shortstack{M360\\Outdoor}} & \multirow{2}{*}{T\&T} & \multirow{2}{*}{DB} \\
Timings in ms &  &  &  &  \\
\midrule
Ours & 7.72 & 10.66 & 7.03 & 5.81 \\
Ours w/o culling & 14.40 & 22.98 & 12.78 & 8.88 \\
Ours w/o hier. sort & 4.11 & 6.29 & 3.69 & 3.47 \\
Ours w/o 3D & 7.64 & 10.30 & 7.52 & 6.32 \\
MCMC & 6.79 & 8.81 & 8.28 & 4.43 \\
\bottomrule
\end{tabular}
    \caption{Average performance timings for different configuration of our method and MCMC.
    As expected, hierarchical sorting introduces the largest performance cost. However, accurate culling compensates for a significant portion of the cost, demonstrating the high efficiency of our 3D culling. Notably, our 3D evaluation is as fast as or even faster than 2D splat approximations.}
    \label{tab:performance_ablation}
\end{table}

%% file: image_metrics.tex
\begin{table*}[ht!]
    \centering
\resizebox{.98\linewidth}{!}{
    \setlength{\tabcolsep}{4pt}
    \begin{tabular}{lcccccccccccc}
    \toprule
     & \multirow{3}{*}{\rotatebox{90}{popping\phantom{x}}} & \multirow{3}{*}{\rotatebox{90}{aliasing\phantom{x}}} & \multirow{3}{*}{\rotatebox{90}{distortion}} \\
    & & & & \multicolumn{3}{c}{Mip-NeRF 360} & \multicolumn{3}{c}{Tanks \& Temples} & \multicolumn{3}{c}{Deep Blending}\\
    \cmidrule(lr){5-7} \cmidrule(lr){8-10} \cmidrule(lr){11-13}
     & & & & PSNR\textsuperscript{$\uparrow$} & SSIM\textsuperscript{$\uparrow$} & LPIPS\textsuperscript{$\downarrow$} & PSNR\textsuperscript{$\uparrow$} & SSIM\textsuperscript{$\uparrow$} & LPIPS\textsuperscript{$\downarrow$} & PSNR\textsuperscript{$\uparrow$} & SSIM\textsuperscript{$\uparrow$} & LPIPS\textsuperscript{$\downarrow$} \\
     \midrule
    3DGS~\cite{kerbl3Dgaussians} & \xmark & \xmark & \xmark & 27.443 & 0.814 & 0.215 & 23.734 & 0.847 & 0.175 & 29.510 & 0.902 & 0.237 \\
    StopThePop~\cite{radl2024stopthepop} &  & \xmark & \xmark & 27.304 & 0.815 & 0.211 & 23.226 & 0.846 & 0.171 & \cellcolor{tab_color!30} 29.929 & \cellcolor{tab_color!10} 0.908 & \cellcolor{tab_color!30} 0.231 \\
    Mip-Splatting~\cite{Yu2023MipSplatting} & \xmark & & \xmark &  \cellcolor{tab_color!00} 27.540 & \cellcolor{tab_color!00} 0.817 & \cellcolor{tab_color!00} 0.216 & \cellcolor{tab_color!10} 23.821 & \cellcolor{tab_color!00} 0.852 & \cellcolor{tab_color!00} 0.176 & \cellcolor{tab_color!00} 29.660 & \cellcolor{tab_color!00} 0.905 & \cellcolor{tab_color!00} 0.243 \\
    MCMC~\cite{kheradmand2024mcmc} & \xmark & \xmark & \xmark & \cellcolor{tab_color!50} 28.027 & \cellcolor{tab_color!30} 0.836 & \cellcolor{tab_color!50} 0.187 & \cellcolor{tab_color!50} 24.642 & \cellcolor{tab_color!50} 0.872 & \cellcolor{tab_color!30} 0.147 & 29.727 & 0.906 & \cellcolor{tab_color!10} 0.233 \\
    Taming 3DGS~\cite{mallick2024taming} & \xmark & \xmark & \xmark & \cellcolor{tab_color!10} 27.826 & \cellcolor{tab_color!10} 0.823 & 0.207 & \cellcolor{tab_color!30} 24.067 & \cellcolor{tab_color!10} 0.855 & \cellcolor{tab_color!10} 0.168 & \cellcolor{tab_color!10} 29.878 & \cellcolor{tab_color!30} 0.910 & 0.235 \\
    Hybrid Transparency\textsuperscript{$\dagger$}~\cite{hahlbohm2024htgs} & $\bm{*}$ & $\bm{*}$ &  & 27.169 & 0.822 & \cellcolor{tab_color!10} 0.195 & - & - & - & - & - & - \\
    Ours &  &  &  & \cellcolor{tab_color!30} 27.835 & \cellcolor{tab_color!50} 0.836 & \cellcolor{tab_color!30} 0.188 & 23.582 & \cellcolor{tab_color!30} 0.867 & \cellcolor{tab_color!50} 0.145 & \cellcolor{tab_color!50} 30.485 & \cellcolor{tab_color!50} 0.913 & \cellcolor{tab_color!50} 0.222 \\
    \bottomrule
    \end{tabular}
}
    \caption{Standard image metrics for our method and related work for in-distribution views. Even though we focus on out-of-distribution effects, our approach matches the state-of-the-art for in-distribution views.
    We also include an overview of the artifacts each method exhibits.
    \textsuperscript{$\bm{*}$}While Hybrid Transparency~\cite{hahlbohm2024htgs} does not suffer from classical popping, they still experience "pop-in" at image borders due to incorrect culling, as well as aliasing due to undersampling of their fixed 3D filter.
    \textsuperscript{$\dagger$}Numbers were re-evaluated on the original evaluation images.}
    \label{tab:evaluation_full}
\end{table*}

%% file: tab_main_comp_ablation.tex
\begin{table*}
    \centering
\resizebox{.95\linewidth}{!}{
    \setlength{\tabcolsep}{4pt}
    \begin{tabular}{lcccccccccccc}
\toprule
& \multirow{3}{*}{\rotatebox{90}{popping\phantom{x}}} & \multirow{3}{*}{\rotatebox{90}{aliasing\phantom{x}}} & \multirow{3}{*}{\rotatebox{90}{distortion}} \\
&&&& \multicolumn{3}{c}{Mip-NeRF 360} & \multicolumn{3}{c}{Tanks \& Temples} & \multicolumn{3}{c}{Deep Blending}\\
\cmidrule(lr){5-7} \cmidrule(lr){8-10} \cmidrule(lr){11-13}
&&&& PSNR\textsuperscript{$\uparrow$} & SSIM\textsuperscript{$\uparrow$} & LPIPS\textsuperscript{$\downarrow$} & PSNR\textsuperscript{$\uparrow$} & SSIM\textsuperscript{$\uparrow$} & LPIPS\textsuperscript{$\downarrow$} & PSNR\textsuperscript{$\uparrow$} & SSIM\textsuperscript{$\uparrow$} & LPIPS\textsuperscript{$\downarrow$}\\
\midrule
Ours \phantom{lorem ipsum dol}  &  &  &  & \cellcolor{tab_color!10} 27.835 & \cellcolor{tab_color!50} 0.836 & \cellcolor{tab_color!30} 0.188 & \cellcolor{tab_color!10} 23.582 & \cellcolor{tab_color!10} 0.867 & \cellcolor{tab_color!30} 0.145 & \cellcolor{tab_color!50} 30.485 & \cellcolor{tab_color!50} 0.913 & \cellcolor{tab_color!50} 0.222 \\
Ours w/o hier. sort             & \xmark &  &  & \cellcolor{tab_color!50} 27.898 & \cellcolor{tab_color!30} 0.836 & \cellcolor{tab_color!00} 0.189 & \cellcolor{tab_color!00} 23.561 & \cellcolor{tab_color!00} 0.865 & \cellcolor{tab_color!10} 0.148 & \cellcolor{tab_color!00} 30.325 & \cellcolor{tab_color!10} 0.912 & \cellcolor{tab_color!00} 0.226 \\
Ours w/o AA                     &  & \xmark &  & \cellcolor{tab_color!00} 27.811 & \cellcolor{tab_color!10} 0.836 & \cellcolor{tab_color!50} 0.185 & \cellcolor{tab_color!30} 23.648 & \cellcolor{tab_color!30} 0.867 & \cellcolor{tab_color!50} 0.141 & \cellcolor{tab_color!10} 30.344 & \cellcolor{tab_color!00} 0.910 & \cellcolor{tab_color!10} 0.224 \\
Ours w/o 3D eval.               & & & \xmark & \cellcolor{tab_color!30} 27.874 & \cellcolor{tab_color!00} 0.836 & \cellcolor{tab_color!10} 0.189 & \cellcolor{tab_color!50} 24.119 & \cellcolor{tab_color!50} 0.869 & \cellcolor{tab_color!00} 0.150 & \cellcolor{tab_color!30} 30.444 & \cellcolor{tab_color!30} 0.912 & \cellcolor{tab_color!30} 0.223 \\
\bottomrule
\end{tabular}
}
    \caption{
Ablation study on the effect of our components for in-distribution views.
Removing individual features may actually increase image metrics, as method can better overfit to the data set views.
See the out-of-distribution evaluation for the benefits of our contributions.
}
    \label{tab:ablation_components}
\end{table*}

%% file: large_fov_tab_singlerowattempt.tex
\begin{table}[!ht]
    \centering
    \footnotesize
    \setlength{\tabcolsep}{2pt}
\begin{tabular}{lrrrrrr}
\toprule
Dataset & \multicolumn{3}{c}{Mip-NeRF 360} & \multicolumn{3}{c}{Tanks \& Temples}\\
\cmidrule(lr){2-4} \cmidrule(lr){5-7}
 & PSNR\textsuperscript{$\uparrow$} & SSIM\textsuperscript{$\uparrow$} & LPIPS\textsuperscript{$\downarrow$} & PSNR\textsuperscript{$\uparrow$} & SSIM\textsuperscript{$\uparrow$} & LPIPS\textsuperscript{$\downarrow$} \\
\midrule
Mip-Splatting   & \cellcolor{tab_color!00} 26.053 & \cellcolor{tab_color!00} 0.784 & \cellcolor{tab_color!00} 0.245 & \cellcolor{tab_color!10} 17.314 & \cellcolor{tab_color!00} 0.737 & \cellcolor{tab_color!00} 0.256 \\
3DGS                & \cellcolor{tab_color!10} 26.820 & \cellcolor{tab_color!10} 0.804 & \cellcolor{tab_color!00} 0.219 & \cellcolor{tab_color!00} 17.112 & \cellcolor{tab_color!10} 0.741 & \cellcolor{tab_color!10} 0.229 \\
StopThePop          & \cellcolor{tab_color!30} 27.040 & \cellcolor{tab_color!30} 0.812 & \cellcolor{tab_color!30} 0.213 & \cellcolor{tab_color!30} 20.241 & \cellcolor{tab_color!30} 0.809 & \cellcolor{tab_color!30} 0.192 \\
Mip-Splatting   & \cellcolor{tab_color!00} 26.053 & \cellcolor{tab_color!00} 0.784 & \cellcolor{tab_color!00} 0.245 & \cellcolor{tab_color!10} 17.314 & \cellcolor{tab_color!00} 0.737 & \cellcolor{tab_color!00} 0.256 \\
MCMC                & \cellcolor{tab_color!00} 23.347 & \cellcolor{tab_color!00} 0.779 & \cellcolor{tab_color!10} 0.213 & \cellcolor{tab_color!00} 14.369 & \cellcolor{tab_color!00} 0.668 & \cellcolor{tab_color!00} 0.296 \\
Taming 3DGS     & \cellcolor{tab_color!00} 23.296 & \cellcolor{tab_color!00} 0.763 & \cellcolor{tab_color!00} 0.231 & \cellcolor{tab_color!00} 11.545 & \cellcolor{tab_color!00} 0.534 & \cellcolor{tab_color!00} 0.448 \\
Ours                & \cellcolor{tab_color!50} 27.836 & \cellcolor{tab_color!50} 0.836 & \cellcolor{tab_color!50} 0.188 & \cellcolor{tab_color!50} 23.583 & \cellcolor{tab_color!50} 0.867 & \cellcolor{tab_color!50} 0.145 \\
\midrule
\end{tabular}
\begin{tabular}{lrrr}
& \multicolumn{3}{c}{Deep Blending}\\
\cmidrule(lr){2-4}
& PSNR\textsuperscript{$\uparrow$} & SSIM\textsuperscript{$\uparrow$} & LPIPS\textsuperscript{$\downarrow$} \\
\midrule
3DGS        & \cellcolor{tab_color!10} 26.192 & \cellcolor{tab_color!10} 0.875 & \cellcolor{tab_color!10} 0.247 \\
StopThePop  & \cellcolor{tab_color!30} 27.553 & \cellcolor{tab_color!30} 0.889 & \cellcolor{tab_color!30} 0.243 \\
Mip-Splatting   & \cellcolor{tab_color!00} 25.592 & \cellcolor{tab_color!00} 0.854 & \cellcolor{tab_color!00} 0.296 \\
MCMC   & \cellcolor{tab_color!00} 18.315 & \cellcolor{tab_color!00} 0.782 & \cellcolor{tab_color!00} 0.355 \\
Taming 3DGS     & \cellcolor{tab_color!00} 20.328 & \cellcolor{tab_color!00} 0.823 & \cellcolor{tab_color!00} 0.282 \\
Ours   & \cellcolor{tab_color!50} 30.488 & \cellcolor{tab_color!50} 0.913 & \cellcolor{tab_color!50} 0.222 \\
\bottomrule
\end{tabular}
    \caption{The large FOV evaluation shows that 3D Gaussian evaluation leads to more faithful reconstruction and rendering.
Compared to related work which relies on evaluating 2D splats on the image plane, our method gracefully retains its image quality in this challenging out-of-distribution rendering scenario.
    }
    \label{tab:large_fov}
\end{table}

%% file: 5_conclusion.tex
\section{Conclusion, Limitations, And Future Work}
\label{sec:conclusion}

In this work, we addressed the limitations of current 3D Gaussian Splatting methods and made several key contributions to enable fast, artifact-free rendering of 3D Gaussians. Our method introduces a novel 3D smoothing filter that effectively removes aliasing artifacts when evaluating Gaussians in 3D, along with stable 3D bounding and culling that performs consistently across various viewing scenarios. We thoroughly evaluated the effectiveness of our components, showing that our method is robust to out-of-distribution camera views while maintaining standard image metrics on par with the current 3DGS state-of-the-art. To our knowledge, we are the only rasterization-based approach capable of delivering artifact-free rendering of 3D Gaussians, with framerates exceeding 100 FPS on consumer-grade hardware.

While achieving more view-consistent results, we observe that standard image metrics do not show significant improvement when evaluation views stay within the training distribution. Although our view-space bounding approach is less tied to the perspective projection, it remains closely tied to the pinhole camera model, which limits its adaptability to other camera models. Furthermore, as our method exhibits stronger view consistency and less room for exploiting view-dependent effects, it would benefit disproportionately from a more expressive view-dependent encoding.

\newpage

\section*{Acknowledgments}

This research was supported by the Austrian Science Fund \textit{FWF} [10.55776/I6663], the German Science Foundation \textit{DFG} [contract 528364066] and the \textit{Alexander von Humboldt Foundation} funded by the German Federal Ministry of Research, Technology and Space. For open access purposes, the author has applied a CC BY public copyright license
to any author-accepted manuscript version arising from this submission.

%% file: 9_appendix.tex
\section{Derivation of Amplitude Scaling Factor}

Let
\[
   \mathbf{d} = \frac{\bm{\upmu} - \mathbf{o}}{\|\bm{\upmu} - \mathbf{o}\|}
\]
be a unit vector in $\mathbb{R}^3$, where $\bm{\upmu}$ is the mean of the Gaussian and $\mathbf{o}$ is the camera position in world space.
We are interested in the area of the Gaussian's intersection with the plane perpendicular to $\mathbf{d}$.

Let
\[
    \mathbf{U} = \begin{pmatrix} \vert & \vert & \vert \\ \mathbf{d} & \mathbf{u}_2 & \mathbf{u}_3 \\ \vert & \vert & \vert\end{pmatrix} \in \mathbb{R}^{3\times3}
\]
be an orthonormal basis with $\mathbf{d}$ as the first basis vector. The orthogonal vectors $\mathbf{u}_2$ and $\mathbf{u}_3$ may be arbitrarily oriented around $\mathbf{d}$, since we are only interested in the size of the area. 
We can perform an orthogonal change of basis on $\mathbf{\Sigma}$:
\[
\mathbf{\Sigma}' = \mathbf{U}^\top \mathbf{\Sigma} \mathbf{U}.
\]
Note that this transformation preserves Eigenvalues, because $\mathbf{U}$ is orthogonal. 
$\mathbf{\Sigma}'$ can be decomposed such that
\[
    \mathbf{\Sigma}' = \begin{pmatrix}
        \sigma_{11} & \bm{\upsigma}_{12}^\top \\ \bm{\upsigma}_{12} & \mathbf{\Sigma}_{\perp},
    \end{pmatrix} \, 
\]
where:
\begin{itemize}
    \item $\sigma_{11} \in \mathbb{R}$ is the $(1, 1)$ entry,
    \item $\bm{\upsigma}_{12} \in \mathbb{R}^2$ is the off-diagonal block,
    \item $\mathbf{\Sigma}_{\perp} \in \mathbb{R}^{2\times 2}$  is an  orthogonal projection of $\mathbf{\Sigma}$ (arbitrarily rotated around $\mathbf{d}$) onto the perpendicular subspace of $\mathbf{d}$.
\end{itemize}

The area of the projected Gaussian is then simply given by the determinant of $\mathbf{\Sigma}_{\perp}$. We can find the determinant by applying the Schur Complement to $\mathbf{\Sigma}'$:
\[
    |\mathbf{\Sigma}'| = |\mathbf{\Sigma}_{\perp}| \cdot \left( \sigma_{11} - \bm{\upsigma}_{12}^\top \mathbf{\Sigma}_{\perp}^{-1} \bm{\upsigma}_{12} \right) \, .
\]
Because of the orthogonal change of basis, $|\mathbf{\Sigma}| = |\mathbf{\Sigma}'|$ and
\begin{equation}
    |\mathbf{\Sigma}_{\perp}| = \frac{1}{\sigma_{11} - \bm{\upsigma}_{12}^\top \mathbf{\Sigma}_{\perp}^{-1} \bm{\upsigma}_{12}} |\mathbf{\Sigma}|.\label{eq:EQ1}
\end{equation}

Using the standard formula for the inverse of a block matrix, we can rewrite the reciprocal of the Schur complement as the $(1, 1)$ entry of $\mathbf{\Sigma}'^{-1}$,
 \begin{equation}
     \label{eq:EQ2}
    \mathbf{e}_1^\top \mathbf{\Sigma}'^{-1} \mathbf{e}_1 = \frac{1}{\sigma_{11} - \bm{\upsigma}_{12}^\top \mathbf{\Sigma}_{22}^{-1} \bm{\upsigma}_{12}} \, 
 \end{equation}
 with $\mathbf{e}_1 = (1, 0, 0)^\top$.
Due to the orthogonality of $\mathbf{U}$,
\begin{align}
\mathbf{e}_1^\top \mathbf{\Sigma}'^{-1} \mathbf{e}_1 &= \mathbf{e}_1^\top (\mathbf{U}^\top \mathbf{\Sigma} \mathbf{U})^{-1} \mathbf{e}_1 \notag\\ &= \mathbf{e}_1^\top \mathbf{U} \mathbf{\Sigma}^{-1} \mathbf{U}^\top \mathbf{e}_1 \notag\\  &=\mathbf{d}^\top  \mathbf{\Sigma}^{-1} \mathbf{d}     \label{eq:EQ3} \text{.}
\end{align}
Combining Eqns. (\ref{eq:EQ1},\ref{eq:EQ2},\ref{eq:EQ3}) results in
\[
 |\mathbf{\Sigma}_{\perp}| = |\mathbf{\Sigma}| \cdot \mathbf{d}^\top \mathbf{\Sigma}^{-1} \mathbf{d} \, ,
\]
and in turn our perpendicular scaling factor is equal to:
 \[
    \sqrt{\frac{|\mathbf{\Sigma}_\perp|}{|\mathbf{\Sigma}_\perp + k \mathbf{I}|}} = \sqrt{\frac{|\mathbf{\Sigma}| \cdot \mathbf{d}^\top \mathbf{\Sigma}^{-1} \mathbf{d}}{|\mathbf{\Sigma}| \cdot \mathbf{d}^\top \left(\mathbf{\Sigma} + k \mathbf{I}\right)^{-1} \mathbf{d}}} \, \text{.}
 \]

\section{Derivation of Bounds}

We perform the plane fitting in view space with planes $\bm{\uppi}_\theta=(\cos(\theta),0,-\sin(\theta),0)^\top, \bm{\uppi}_\phi=(0,\cos(\phi),-\sin(\phi),0)^\top$, and their transformed counterparts in Gaussian space $\bm{\uppi'}_\theta, \bm{\uppi'}_\phi$:
\begin{align}
    & \bm{\uppi'}_\theta=\mathbf{T}_\text{view}^\top \bm{\uppi}_\theta = \cos(\theta) \mathbf{T}_{(\text{view},1)} - \sin(\theta) \mathbf{T}_{(\text{view},3)}, \\
    & \bm{\uppi'}_\phi=\mathbf{T}_\text{view}^\top \bm{\uppi}_\phi = \cos(\phi) \mathbf{T}_{(\text{view},2)} - \sin(\phi) \mathbf{T}_{(\text{view},3)},
\end{align}
with $\mathbf{T}_\text{view}=\mathbf{V}\mathbf{T}$ being the transformation matrix from Gaussian space to view space via view-matrix $\mathbf{V}$, and $\mathbf{T}_{(\text{view},i)}$ denoting the $i$-th row of $\mathbf{T}_\text{view}$. For simplicity, we will refer to $\mathbf{T}_{(\text{view},i)}$ as $\mathbf{T}_i$ in the following derivation. \\

Following \citet{botsch2006quadraticsurfaces}, the touching condition to the cutoff ellipsoid in Gaussian space for these planes is
\begin{align}
    \bm{\uppi'}^\top \mathbf{Q} \bm{\uppi'} = 0
\end{align}
with $\mathbf{Q}\in\mathbb{R}^{4\times4}$ being a diagonal matrix, which is defined as $\mathbf{Q} = \text{diag}(\bm{t}), \bm{t}=(\tau_\rho,\tau_\rho,\tau_\rho,-1)^\top$. For $\bm{\uppi'}_\theta$, this simplifies to
{\scriptsize
    \begin{align*}
         &(\cos(\theta) \mathbf{T}_1 - \sin(\theta)\mathbf{T}_3)^\top \mathbf{Q} (\cos(\theta) \mathbf{T}_1 - \sin(\theta)\mathbf{T}_3)\\
         = &\cos(\theta)^2 \mathbf{T}_1^\top \mathbf{Q} \mathbf{T}_1 - 2\sin(\theta) \cos(\theta) \mathbf{T}_1^\top \mathbf{Q} \mathbf{T}_3 + \sin(\theta)^2 \mathbf{T}_3^\top \mathbf{Q} \mathbf{T}_3 \\
         = &\tan(\theta)^2 \mathbf{T}_1^\top \mathbf{Q} \mathbf{T}_1 - 2\tan(\theta) \mathbf{T}_1^\top \mathbf{Q} \mathbf{T}_3 + \mathbf{T}_3^\top \mathbf{Q} \mathbf{T}_3 \\
         = &\tan(\theta)^2 \langle \bm{t}, \mathbf{T}_1 \odot \mathbf{T}_1 \rangle - 2\tan(\theta) \langle \bm{t}, \mathbf{T}_1 \odot \mathbf{T}_3 \rangle + \langle \bm{t}, \mathbf{T}_3 \odot \mathbf{T}_3 \rangle.
    \end{align*}
}%
By solving this quadratic equation w.r.t. $\tan(\theta)$ (and similarly $\tan(\phi)$), we find solutions for $\theta,\phi$: 
\begin{align}
    \theta_{1,2}=\tan^{-1} \left( \frac{s_{1,3} \pm \sqrt{s_{1,3}^2 - s_{1,1}s_{3,3}}}{s_{3,3}} \right), \\
    \phi_{1,2}=\tan^{-1} \left( \frac{s_{2,3} \pm \sqrt{s_{2,3}^2 - s_{2,2}s_{3,3}}}{s_{3,3}} \right),
\end{align}
with $s_{i,j}=\langle \mathbf{t}, \mathbf{T}_i \odot \mathbf{T}_j\rangle$.
This closely relates to the bounds computed by \citet{hahlbohm2024htgs}, but allows for the analysis and bounding of angles before transforming them to the screen, instead of directly receiving screen bounds.

\section{Multi-Resolution Evaluation Images}

\begin{figure*}
    \centering
    \includegraphics[width=0.7\linewidth]{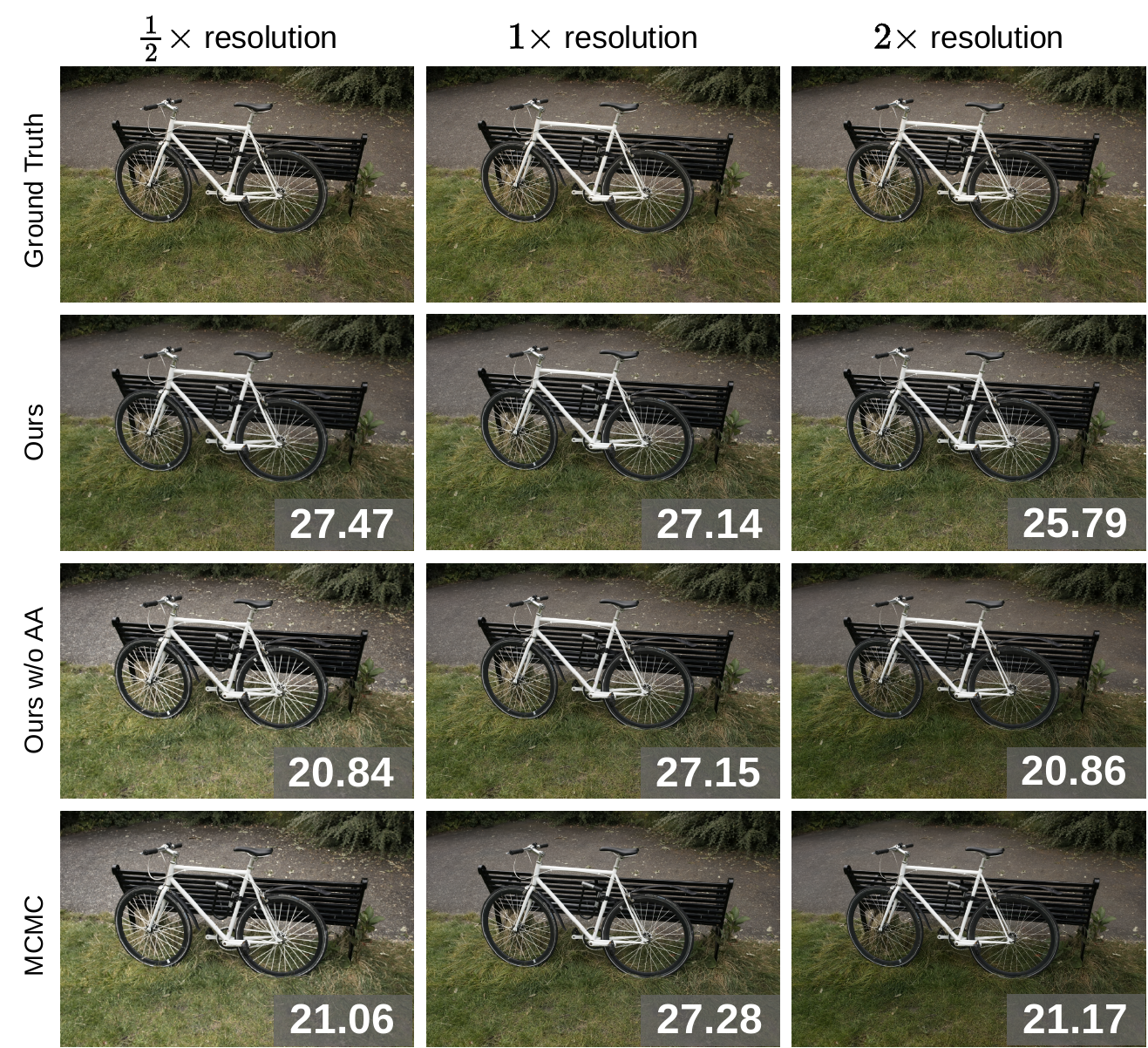}
    \caption{A single view of our multi-resolution evaluation on the Mip-Nerf~360 bicycle scene, with inset PSNR values.}
    \label{fig:suppl_multiresolution_bicycle}
\end{figure*}

We show an example view of our multi-resolution evaluation in \cref{fig:suppl_multiresolution_bicycle}. While Ours is able to retain good image quality at all resolution levels, and correctly dilates and smooths content.
In contrast, MCMC and our method without the anti-aliasing 3D smoothing filter exhibit considerable aliasing: content becomes too thick on lower resolution and too thin on higher resolution.
This also shows in the inset PSNR values.

\section{Per Scene Image Metrics}

We provide per-scene image metrics for all our evaluated scenes in \cref{tab:suppl_perscene_full}.

\begin{table*}[!ht]
    \centering
        \caption{Per-scene image metrics for all methods on all evaluated scenes.}\setlength{\tabcolsep}{4pt}
    \scriptsize
    \setlength{\tabcolsep}{2pt}
    \begin{tabular}{lccccccccccccc}
\toprule
Dataset & \multicolumn{5}{c}{Mip-NeRF 360 Outdoor} &\multicolumn{4}{c}{Mip-NeRF 360 Indoor} & \multicolumn{2}{c}{Deep Blending} & \multicolumn{2}{c}{Tanks \& Temples}\\
Scene & Bicycle & Flowers & Garden & Stump & Treehill & Bonsai & Counter & Kitchen & Room & DrJ & Playroom & Train & Truck\\\midrule
& \\[-2.4ex]
 & \multicolumn{13}{c}{PSNR\textsuperscript{$\uparrow$}} \\\cmidrule(lr){2-14}
3DGS                & \cellcolor{tab_color!0} 25.19 & \cellcolor{tab_color!0} 21.53 & \cellcolor{tab_color!0} 27.30 & \cellcolor{tab_color!0} 26.62 & \cellcolor{tab_color!0} 22.46 & \cellcolor{tab_color!0} 32.11 & \cellcolor{tab_color!0} 28.97 & \cellcolor{tab_color!0} 31.33 & \cellcolor{tab_color!0} 31.48 & \cellcolor{tab_color!0} 29.05 & \cellcolor{tab_color!0} 29.97 & \cellcolor{tab_color!0} 22.05 & \cellcolor{tab_color!0} 25.41 \\
StopThePop          & \cellcolor{tab_color!0} 25.22 & \cellcolor{tab_color!0} 21.54 & \cellcolor{tab_color!0} 27.23 & \cellcolor{tab_color!0} 26.70 & \cellcolor{tab_color!0} 22.44 & \cellcolor{tab_color!0} 31.98 & \cellcolor{tab_color!0} 28.60 & \cellcolor{tab_color!0} 31.18 & \cellcolor{tab_color!0} 30.84 & \cellcolor{tab_color!0} 29.45 & \cellcolor{tab_color!30} 30.40 & \cellcolor{tab_color!0} 21.49 & \cellcolor{tab_color!0} 24.96 \\
Mip-Splatting       & \cellcolor{tab_color!0} 25.32 & \cellcolor{tab_color!0} 21.64 & \cellcolor{tab_color!0} 27.48 & \cellcolor{tab_color!0} 26.58 & \cellcolor{tab_color!0} 22.58 & \cellcolor{tab_color!0} 32.13 & \cellcolor{tab_color!0} 29.00 & \cellcolor{tab_color!0} 31.34 & \cellcolor{tab_color!10} 31.78 & \cellcolor{tab_color!0} 29.15 & \cellcolor{tab_color!0} 30.17 & \cellcolor{tab_color!10} 22.16 & \cellcolor{tab_color!10} 25.48 \\
MCMC           & \cellcolor{tab_color!30} 25.69 & \cellcolor{tab_color!30} 22.01 & \cellcolor{tab_color!50} 27.87 & \cellcolor{tab_color!50} 27.36 & \cellcolor{tab_color!30} 22.94 & \cellcolor{tab_color!50} 32.65 & \cellcolor{tab_color!50} 29.38 & \cellcolor{tab_color!50} 32.09 & \cellcolor{tab_color!50} 32.25 & \cellcolor{tab_color!30} 29.52 & \cellcolor{tab_color!0} 29.93 & \cellcolor{tab_color!50} 22.83 & \cellcolor{tab_color!50} 26.45 \\
Ours           & \cellcolor{tab_color!50} 25.74 & \cellcolor{tab_color!50} 22.13 & \cellcolor{tab_color!10} 27.50 & \cellcolor{tab_color!30} 27.24 & \cellcolor{tab_color!50} 23.03 & \cellcolor{tab_color!10} 32.32 & \cellcolor{tab_color!30} 29.24 & \cellcolor{tab_color!30} 31.88 & \cellcolor{tab_color!0} 31.45 & \cellcolor{tab_color!50} 29.90 & \cellcolor{tab_color!50} 31.07 & \cellcolor{tab_color!0} 21.73 & \cellcolor{tab_color!0} 25.43 \\
Taming 3DGS         & \cellcolor{tab_color!10} 25.47 & \cellcolor{tab_color!10} 21.87 & \cellcolor{tab_color!30} 27.76 & \cellcolor{tab_color!10} 27.05 & \cellcolor{tab_color!10} 22.91 & \cellcolor{tab_color!30} 32.47 & \cellcolor{tab_color!10} 29.06 & \cellcolor{tab_color!10} 31.76 & \cellcolor{tab_color!30} 32.09 & \cellcolor{tab_color!10} 29.51 & \cellcolor{tab_color!10} 30.24 & \cellcolor{tab_color!30} 22.25 & \cellcolor{tab_color!30} 25.88 \\
Hybrid Transparency & \cellcolor{tab_color!0} 25.31 & \cellcolor{tab_color!0} 21.35 & \cellcolor{tab_color!0} 27.22 & \cellcolor{tab_color!0} 26.85 & \cellcolor{tab_color!0} 22.37 & \cellcolor{tab_color!0} 31.55 & \cellcolor{tab_color!0} 28.40 & \cellcolor{tab_color!0} 31.02 & \cellcolor{tab_color!0} 30.45 & \cellcolor{tab_color!0} - & \cellcolor{tab_color!00} - & \cellcolor{tab_color!0} - & \cellcolor{tab_color!0} - \\
\midrule
& \\[-2.4ex]
 & \multicolumn{13}{c}{SSIM\textsuperscript{$\uparrow$}} \\\cmidrule(lr){2-14}
3DGS & \cellcolor{tab_color!0} 0.764 & \cellcolor{tab_color!0} 0.605 & \cellcolor{tab_color!0} 0.864 & \cellcolor{tab_color!0} 0.772 & \cellcolor{tab_color!0} 0.633 & \cellcolor{tab_color!0} 0.941 & \cellcolor{tab_color!0} 0.907 & \cellcolor{tab_color!0} 0.926 & \cellcolor{tab_color!0} 0.919 & \cellcolor{tab_color!0} 0.900 & \cellcolor{tab_color!0} 0.905 & \cellcolor{tab_color!0} 0.814 & \cellcolor{tab_color!0} 0.880 \\
StopThePop & \cellcolor{tab_color!0} 0.768 & \cellcolor{tab_color!0} 0.605 & \cellcolor{tab_color!0} 0.864 & \cellcolor{tab_color!0} 0.776 & \cellcolor{tab_color!0} 0.635 & \cellcolor{tab_color!0} 0.941 & \cellcolor{tab_color!0} 0.905 & \cellcolor{tab_color!0} 0.926 & \cellcolor{tab_color!0} 0.918 & \cellcolor{tab_color!10} 0.906 & \cellcolor{tab_color!10} 0.910 & \cellcolor{tab_color!0} 0.810 & \cellcolor{tab_color!0} 0.882 \\
Mip-Splatting & \cellcolor{tab_color!0} 0.768 & \cellcolor{tab_color!0} 0.608 & \cellcolor{tab_color!0} 0.869 & \cellcolor{tab_color!0} 0.773 & \cellcolor{tab_color!0} 0.638 & \cellcolor{tab_color!0} 0.942 & \cellcolor{tab_color!0} 0.909 & \cellcolor{tab_color!0} 0.928 & \cellcolor{tab_color!0} 0.920 & \cellcolor{tab_color!0} 0.902 & \cellcolor{tab_color!0} 0.909 & \cellcolor{tab_color!0} 0.818 & \cellcolor{tab_color!0} 0.886 \\
MCMC & \cellcolor{tab_color!30} 0.799 & \cellcolor{tab_color!30} 0.645 & \cellcolor{tab_color!50} 0.878 & \cellcolor{tab_color!30} 0.811 & \cellcolor{tab_color!30} 0.659 & \cellcolor{tab_color!50} 0.948 & \cellcolor{tab_color!50} 0.917 & \cellcolor{tab_color!50} 0.934 & \cellcolor{tab_color!50} 0.930 & \cellcolor{tab_color!0} 0.904 & \cellcolor{tab_color!0} 0.908 & \cellcolor{tab_color!50} 0.843 & \cellcolor{tab_color!30} 0.900 \\
Ours & \cellcolor{tab_color!50} 0.801 & \cellcolor{tab_color!50} 0.648 & \cellcolor{tab_color!30} 0.876 & \cellcolor{tab_color!50} 0.813 & \cellcolor{tab_color!50} 0.662 & \cellcolor{tab_color!30} 0.948 & \cellcolor{tab_color!30} 0.917 & \cellcolor{tab_color!30} 0.934 & \cellcolor{tab_color!30} 0.928 & \cellcolor{tab_color!50} 0.910 & \cellcolor{tab_color!50} 0.916 & \cellcolor{tab_color!30} 0.833 & \cellcolor{tab_color!50} 0.900 \\
Taming 3DGS & \cellcolor{tab_color!0} 0.780 & \cellcolor{tab_color!0} 0.614 & \cellcolor{tab_color!10} 0.873 & \cellcolor{tab_color!0} 0.788 & \cellcolor{tab_color!10} 0.646 & \cellcolor{tab_color!10} 0.944 & \cellcolor{tab_color!10} 0.910 & \cellcolor{tab_color!10} 0.931 & \cellcolor{tab_color!10} 0.924 & \cellcolor{tab_color!30} 0.909 & \cellcolor{tab_color!30} 0.911 & \cellcolor{tab_color!10} 0.819 & \cellcolor{tab_color!10} 0.892 \\
Hybrid Transparency & \cellcolor{tab_color!10} 0.785 & \cellcolor{tab_color!10} 0.631 & \cellcolor{tab_color!0} 0.867 & \cellcolor{tab_color!10} 0.793 & \cellcolor{tab_color!0} 0.639 & \cellcolor{tab_color!0} 0.941 & \cellcolor{tab_color!0} 0.902 & \cellcolor{tab_color!0} 0.923 & \cellcolor{tab_color!0} 0.920 & \cellcolor{tab_color!0} - & \cellcolor{tab_color!0} - & \cellcolor{tab_color!0} - & \cellcolor{tab_color!0} - \\ 
\midrule
& \\[-2.4ex]
 & \multicolumn{13}{c}{LPIPS\textsuperscript{$\downarrow$}} \\\cmidrule(lr){2-14}
3DGS & \cellcolor{tab_color!0} 0.210 & \cellcolor{tab_color!0} 0.335 & \cellcolor{tab_color!0} 0.107 & \cellcolor{tab_color!0} 0.214 & \cellcolor{tab_color!0} 0.326 & \cellcolor{tab_color!0} 0.200 & \cellcolor{tab_color!0} 0.198 & \cellcolor{tab_color!0} 0.125 & \cellcolor{tab_color!0} 0.216 & \cellcolor{tab_color!0} 0.240 & \cellcolor{tab_color!0} 0.234 & \cellcolor{tab_color!0} 0.205 & \cellcolor{tab_color!0} 0.144 \\
StopThePop & \cellcolor{tab_color!0} 0.204 & \cellcolor{tab_color!0} 0.332 & \cellcolor{tab_color!0} 0.106 & \cellcolor{tab_color!0} 0.208 & \cellcolor{tab_color!0} 0.316 & \cellcolor{tab_color!0} 0.199 & \cellcolor{tab_color!0} 0.197 & \cellcolor{tab_color!0} 0.125 & \cellcolor{tab_color!0} 0.214 & \cellcolor{tab_color!30} 0.231 & \cellcolor{tab_color!30} 0.231 & \cellcolor{tab_color!10} 0.202 & \cellcolor{tab_color!0} 0.140 \\
Mip-Splatting & \cellcolor{tab_color!0} 0.212 & \cellcolor{tab_color!0} 0.339 & \cellcolor{tab_color!0} 0.108 & \cellcolor{tab_color!0} 0.216 & \cellcolor{tab_color!0} 0.326 & \cellcolor{tab_color!0} 0.204 & \cellcolor{tab_color!0} 0.200 & \cellcolor{tab_color!0} 0.126 & \cellcolor{tab_color!0} 0.218 & \cellcolor{tab_color!0} 0.243 & \cellcolor{tab_color!0} 0.243 & \cellcolor{tab_color!0} 0.205 & \cellcolor{tab_color!0} 0.147 \\
MCMC & \cellcolor{tab_color!50} 0.168 & \cellcolor{tab_color!30} 0.284 & \cellcolor{tab_color!50} 0.094 & \cellcolor{tab_color!50} 0.171 & \cellcolor{tab_color!50} 0.272 & \cellcolor{tab_color!10} 0.191 & \cellcolor{tab_color!30} 0.185 & \cellcolor{tab_color!30} 0.121 & \cellcolor{tab_color!30} 0.198 & \cellcolor{tab_color!10} 0.234 & \cellcolor{tab_color!10} 0.233 & \cellcolor{tab_color!50} 0.183 & \cellcolor{tab_color!30} 0.112 \\
Ours & \cellcolor{tab_color!30} 0.171 & \cellcolor{tab_color!10} 0.285 & \cellcolor{tab_color!30} 0.099 & \cellcolor{tab_color!30} 0.171 & \cellcolor{tab_color!10} 0.275 & \cellcolor{tab_color!30} 0.189 & \cellcolor{tab_color!50} 0.183 & \cellcolor{tab_color!50} 0.121 & \cellcolor{tab_color!50} 0.197 & \cellcolor{tab_color!50} 0.223 & \cellcolor{tab_color!50} 0.220 & \cellcolor{tab_color!30} 0.184 & \cellcolor{tab_color!50} 0.106 \\
Taming 3DGS & \cellcolor{tab_color!0} 0.192 & \cellcolor{tab_color!0} 0.332 & \cellcolor{tab_color!10} 0.100 & \cellcolor{tab_color!0} 0.196 & \cellcolor{tab_color!0} 0.313 & \cellcolor{tab_color!0} 0.201 & \cellcolor{tab_color!0} 0.198 & \cellcolor{tab_color!10} 0.122 & \cellcolor{tab_color!0} 0.210 & \cellcolor{tab_color!0} 0.234 & \cellcolor{tab_color!0} 0.235 & \cellcolor{tab_color!0} 0.208 & \cellcolor{tab_color!10} 0.128 \\
Hybrid Transparency & \cellcolor{tab_color!10} 0.178 & \cellcolor{tab_color!50} 0.282 & \cellcolor{tab_color!0} 0.106 & \cellcolor{tab_color!10} 0.193 & \cellcolor{tab_color!30} 0.275 & \cellcolor{tab_color!50} 0.189 & \cellcolor{tab_color!10} 0.197 & \cellcolor{tab_color!0} 0.128 & \cellcolor{tab_color!10} 0.205 & \cellcolor{tab_color!0} - & \cellcolor{tab_color!0} - & \cellcolor{tab_color!0} - & \cellcolor{tab_color!0} - \\
\bottomrule
\end{tabular}

    \label{tab:suppl_perscene_full}
\end{table*}